\documentclass[natbib]{svjour3}             
\smartqed  
\usepackage{graphicx}
\def\ie{{ i.e.,\ }}
\newcommand{\BellAmp}{N_B}
\newcommand{\be}{\begin{eqnarray}}
\newcommand{\ee}{\end{eqnarray}}
\newcommand{\lsim}{\;\raise0.3ex\hbox{$<$\kern-0.75em\raise-1.1ex\hbox{$\sim$}}\;}
\newcommand{\gsim}{\;\raise0.3ex\hbox{$>$\kern-0.75em\raise-1.1ex\hbox{$\sim$}}\;}

\newcommand{\nCR}{n_\mathrm{cr}}

\newcommand{\jCRmeanVec}{\mathbf{j}^\mathrm{cr}}

\newcommand{\bSqMean}{\langle b^{2}_{B}\rangle}
\newcommand{\taucor}{\tau_\mathrm{cor}}
\newcommand{\alpt}{\alpha_t}

\newcommand{\meanEMF}{\overline{\mathbf{\mathcal{E}}}}
\newcommand{\uu}{\mathbf{u}}
\newcommand{\bb}{\mathbf{b}}
\newcommand{\meanemf}{\overline{\mathcal{E}}}
\newcommand{\meanB}{\overline{B}}
\newcommand{\jj}{\mathbf{j}}
\newcommand{\BB}{\mathbf{B}}
\newcommand{\meanA}{\overline{A}}
\newcommand{\meanBB}{\overline{\mathbf{B}}}
\newcommand{\vAz}{v_{A0}}
\newcommand{\vA}{v_{a}}
\newcommand{\Lu}{\textup{Lu}}
\newcommand{\OO}{\mathbf{\Omega}}
\newcommand{\grav}{\mathbf{g}}
\newcommand{\etaT}{\eta_{T}}
\newcommand{\etat}{\eta_{t}}
\newcommand{\nuM}{\nu_{M}}
\newcommand{\nab}{\nabla}
\newcommand{\ii}{\textup{i}}
\newcommand{\kk}{\mathbf{k}}
\newcommand{\xx}{\mathbf{x}}
\newcommand{\zzz}{\hat{\mathbf{z}}}
\newcommand{\meanJ}{\overline{J}}
\newcommand{\Eq}[1]{Eq.(\ref{#1})}
\newcommand{\meanJJ}{\overline{\mathbf{J}}}
\newcommand{\meanKK}{\overline{\mathbf{K}}}
\newcommand{\meanK}{\overline{K}}
\newcommand{\Fig}[1]{Figure \ref{#1}}
\newcommand{\Eqs}[2]{Eqs.(\ref{#1}) and (\ref{#2})}
\def\lsim{\;\raise0.3ex\hbox{$<$\kern-0.75em\raise-1.1ex\hbox{$\sim$}}\;}
\def\gsim{\;\raise0.3ex\hbox{$>$\kern-0.75em\raise-1.1ex\hbox{$\sim$}}\;}

\usepackage{color}

\begin{document}

\title{
}
\title{Microphysics of cosmic ray driven plasma instabilities}



\author{A.M. Bykov \and A. Brandenburg
 \and  \\ M.A. Malkov \and S.M. Osipov
}

\institute{A.M.~Bykov \at A.F.~Ioffe Institute for Physics and
Technology, 194021, St.Petersburg, Russia \\and St.Petersburg State
Politechnical University, \email{byk@astro.ioffe.ru}\and
A.~Brandenburg \at Nordita, Royal Institute of Technology and
Stockholm University, Roslagstullsbacken 23, 10691 Stockholm,
Sweden; and Department of Astronomy, Stockholm University, SE 10691
Stockholm, Sweden, \email{brandenb@nordita.org} \and M.A.~Malkov \at
University of California, San Diego, La Jolla, California 92093, USA
\email{mmalkov@ucsd.edu} \and S.M.~Osipov \at A.F.~Ioffe Institute
for Physics and Technology, 194021, St.Petersburg, Russia
\email{osm2004@mail.ru} }

\date{Received: date / Accepted: date}

\maketitle

\begin{abstract}
Energetic nonthermal particles (cosmic rays, CRs) are accelerated in
supernova remnants, relativistic jets and other astrophysical
objects. The CR energy density is typically comparable with that of
the thermal components and magnetic fields. In this review we
discuss mechanisms of magnetic field amplification due to
instabilities induced by CRs. We derive CR kinetic and
magnetohydrodynamic equations that govern cosmic plasma systems
comprising the thermal background plasma, comic rays and fluctuating
magnetic fields to study CR-driven instabilities. Both resonant and
non-resonant instabilities are reviewed, including the Bell
short-wavelength instability,
 and the firehose instability. Special attention is paid
to the longwavelength instabilities driven by the CR current and
pressure gradient. The helicity production by the CR current-driven
instabilities is discussed in connection with the dynamo mechanisms
of cosmic magnetic field amplification. 
\end{abstract}

\maketitle

\section{Introduction}
Acceleration of cosmic rays (CRs) in the Galaxy by the first order
Fermi mechanism is believed to be very efficient. Most of the
theoretical studies of shock acceleration agree on its potential to
convert, under \emph{favorable conditions,} 50\% or more of shock
mechanical energy into the CR energy. Observational estimates of the
supernova remnant (SNR) shock power require, \emph{on the average},
a 15-30\%\emph{ }conversion efficiency to maintain the observed CR
energy against losses from the Galaxy \citep[see,
e.g.,][]{berea90,DruMarkVlk89}. However, this acceleration mechanism
is fast enough only if it is self-sustained; accelerated particles
must be scattered across the shock at an enhanced rate (to gain
energy rapidly) by magnetic irregularities amplified by the
particles themselves. Relying on the background magnetic
irregularities (interstellar medium {[}ISM{]} turbulence) would
result only in a very slow acceleration.

Fortunately, freshly accelerated CRs indeed comprise enough free
energy to drive plasma instabilities thus bootstrapping their own
acceleration \citep[see, e.g.,][]{zweibel79}. While they are
accumulated in a relatively thin layer near a shock front, their
pressure gradient is built up. Furthermore, they stream through the
inflowing plasma so that their pitch-angle distribution is
anisotropic. They also provide an electric current and induce a
return current in the upstream plasma.

Instabilities driven by the above sources of free energy may loosely
be categorized as follows. First, an ion-cyclotron type, resonant
instability (driven by the CR anisotropy) amplifies Alfven and
magnetosonic waves, with no major changes to their dispersive
properties and the macroscopic state of the medium near the shock.
However, the amplified waves make the CR pressure and current to
build-up rapidly through an enhanced CR scattering and energy gain.
Second, there is a non-resonant firehose type instability driven by
the CR pressure anisotropy. In contrast to the resonant instability,
the firehose instability changes the Alfven wave dispersive
properties by making the growing mode aperiodic. So does the current
driven non-resonant instability. The renewed interest to this
instability has been sparked by \cite{Bell04}, who revealed its
potential to strongly amplify the background magnetic field. Indeed,
a formal analytic solution in which the instability driver is
balanced by the nonlinearity indicates that the instability
saturates only at very high amplitudes, $\delta B\gg B_{0}$
\citep[see
e.g.][]{BelLuc01,bell05,marcowithea06,CBAV2008,vbe09,MDS_Solitons12}.
Finally, the CR pressure gradient in the shock precursor drives
acoustic perturbations. All these instabilities should be treated on
a unified basis, as they are driven by the anisotropic inhomogeneous
CR plasma component near a shock front. An attempt of such treatment
is presented below. However a complete nonlinear study of these
phenomena is a formidable task, yet to be accomplished.

While the above instabilities, clearly associated with collisionless
shocks, will be central to the present review, CRs are also known to
drive instabilities crucial to their confinement regardless of the
way they are accelerated. For example, a sufficiently dense CR cloud
released into the ISM will drive Alfven waves which, in turn, will
scatter the CRs, thus delaying their escape \citep[see
e.g.][]{PtuskinNLDIFF08,ohiraea11,Malkov_etal_escape12,YanLazarianEscape12}. Moving further
out to the CR confinement in the galaxy, the so-called Parker
instability is known to be important, in addition to the Alfven wave
self-generation by escaping CRs.

The diffusive shock acceleration (DSA) mechanism is based on
repeated shock crossings with a $\sim u_{\rm s}/c$ particle energy
gain per cycle
\citep[see][]{krym77,bell78,Blandford87,berkrym88,je91}. While doing
so, particles diffusively escape from the shock up to a distance
$L_{p}\sim\kappa\left(p\right)/u_{s}$. Here $\kappa$ is the momentum
dependent diffusion coefficient and $u_{\rm s}$ is the shock
velocity. One should expect then an extended ($\sim L_{p}$) shock
precursor populated by accelerated protons and electrons so that
synchrotron radiating electrons may make it visible. High-resolution
$X$-ray observations have revealed  thin X-ray synchrotron filaments
and fast evolving clumps in synchrotron emitting supernova shells.
The filaments are much thinner than $L_{p}$ because the TeV regime
electrons  are confined in a narrow layer around the shock. Most
likely they are limited by fast synchrotron cooling due to the X-ray
emission in a highly amplified magnetic field \citep[see for
review][]{CassamChenHughes07,reyn08,vink12,helderea12}. The
synchrotron emission clumps with a year time scale variability
observed with {\sl Chandra} observatory by \citet[][]{uchiyamaea07}
can be associated with strong intermittency of the amplified
magnetic fields \citep{bue08}. Moreover, a quasi-regular set of
strips of synchrotron emission resolved with {\sl Chandra} in
Tycho's SNR by \citet{Eriksen11} potentially can be used to study a
specific angular dependence and the spectral properties of nonlinear
mechanisms of magnetic field amplification by CR-driven
instabilities \citep[][]{Tycho_stripes11}.

 According to the widely accepted view, the particle
diffusion coefficient $\kappa$ should be close to the Bohm value,
$\kappa\sim cr_{g}\left(p\right)/3$, which requires strong magnetic
fluctuations $\delta B_{k}\sim B_{0}$ at the resonant scale
$k\sim1/r_{g}\left(p\right)$. The high level of fluctuations is
achieved through one of the instabilities driven by accelerated
particles. A number of CR driven instabilities have been suggested
to generate magnetic field fluctuations. The first one is the well
known ion cyclotron resonant instability of a slightly anisotropic
(in pitch angle) CR distribution \citep[see
e.g.][]{SagdShafr61,zweibel79,schlikeiser02,amato11}. The free
energy source of this instability is potentially sufficient to
generate magnetic field fluctuations needed to scatter CRs ahead of
the shock \citep[see e.g.][]{bell78,McKVlk82}.

\begin{equation}
\left(\delta B/B_{0}\right)^{2}\sim M_{A}P^{cr}/\rho u_{\rm s}^{2}.\label{delB}
\end{equation}
where $M_{A}\gg1$ is the Alfvenic Mach number, $P^{cr}$ is the CR
pressure, $\rho$ is the gas density and $u_{\rm s}$ is the shock
velocity. However, the actual turbulence level was shown to remain
moderate, $\delta B\sim B_{0}$  as this is a resonant kinetic
instability that is usually suppressed by a quasilinear
isotropisation or particle trapping effects easily \citep[see
e.g.][]{McKVlk82,AchtBland86,zweibel03}.

The second instability, is a nonresonant instability driven by the
CR current. The advantage of this instability seems to be twofold.
First, it cannot be stabilized by the quasilinear deformation of the
CR distribution function since in the upstream plasma frame the
driving CR current persists, once the CR cloud is at rest in the
shock frame. Second, it generates a broad spectrum of waves, and the
longest ones were claimed to be stabilized only at the level $\delta
B\gg B_{0}$, due to the lack of efficient stabilization mechanism at
such scales \citep[see e.g.][]{Bell04}. Within the context of the CR
acceleration, this instability was  studied by \citet{achterberg83}
\citep[see also][]{ShapiroQuest98GeoRL},  but the fast regime of the
nonresonant instability was found by \citet{BelLuc01} and
\citet{Bell04}, and therefore the instability is often referred to
as Bell's instability. \citet{Bell04} pointed out that in the
instability is driven by a fixed CR return current through the
Ampere force $\mathbf{j}_{cr}\times\mathbf{B}$. It should be noted,
however, that the dissipation of the return current due to the
anomalous resistivity still needs to be addressed. The effect of a
finite plasma temperature on the instability was studied by
\citet[][]{ze10}. Actually, as we will show below, both the resonant
and the Bell instabilities are interconnected, they are driven by
the CR drift relative the background plasma. Moreover, in the case
of the modes propagating along the mean magnetic fields the two
instabilities are simultaneously influencing the same modes. The
dispersion relations of the modes are strongly influenced by the
presence of the CR current are markedly different from the standard
MHD modes. The dispersion relations of the modes strongly influenced
by the presence of the CR current are markedly different from the
standard MHD modes. The dispersion relation in the longwavelength
regime (where the mode wavelengths are larger than the bulk CR
gyroradii) can be also strongly modified by the ponderomotive forces
induced by Bell's turbulence. The longwavelength instability has two
regimes \citep[][]{Bykov11,ber12}. The first regime is prominent in
the intermediate range where the mode wavelength is above the CR
gyroradii but below the CR mean free path. It is discussed in
\S\ref{dynamo} and is associated with a dynamo type instability
driven by the nonzero helicity, which is, in turn, produced by the
short scale CR-driven turbulence. The intermediate wavenumber range
is rather narrow in the case of the Bohm-type CR diffusion. The
modes with wavelengths larger than the CR mean free path are subject
of non-resonant long-wavelength instability caused by the
ponderomotive force acting on the background plasma that is induced
by Bell's turbulence. We discuss the long wavelength instability
below in \S\ref{LW}.

The third instability is an acoustic instability (also known as
Drury's instability) driven by the pressure gradient of accelerated
CRs upstream \citep{dorfi85,DruryFal86,DruryDrInst12,schureea12}.
The pressure gradient is clearly a viable source of free energy for
the instability. So, among the macroscopic quantities varying across
a strong shock, the pressure jump is the most pronounced one in that
it does not saturate with the Mach number, unlike the density or
velocity jumps.

The acoustic instability has received somewhat less attention than
the first two. Moreover, in many numerical studies of the CR shock
acceleration, special care is taken to suppress it. The suppression
is achieved by using the fact that a change of stability occurs at
that point in the flow where
$\partial\ln\kappa/\partial\ln\rho\simeq-1$ (for both stable and
unstable wave propagation directions, of course, if such point
exists at all). Here $\rho$ is the gas density. Namely, one requires
this condition to hold identically all across the shock precursor,
\ie where the CR pressure gradient $\nabla P^{cr}\neq0$. Not only is
this requirement difficult to justify physically, but, more
importantly, an \emph{artificial} suppression of the instability
eliminates its \emph{genuine }macroscopic and microscopic
consequences, as briefly discussed below.

Among the macroscopic consequences  an important one is the
vorticity generation through the baroclinic effect (missalignment of
the density and pressure gradients $\nabla\rho\times\nabla P\neq0$,
e.g. \cite{rkj93,KulsrudMF97}). Here $\nabla P$ may be associated
with a quasi-constant macroscopic CR-gas pressure gradient $\nabla
P^{cr}$, generally directed along the shock normal. Variations of
$\nabla\rho$ are locally decoupled from $P^{cr}$, unlike in the
situation in a gas with a conventional equation of state where
$P=P\left(\rho\right)$ and where the baroclinic term vanishes. The
vorticity generation obviously results (just through the frozen in
condition) in magnetic field generation, so that the field can be
amplified by the CR pressure gradient. More importantly, this
process amplifies the \emph{large scale field}, required for
acceleration of \emph{high energy particles}. Furthermore, the
amplification takes place well ahead of the gaseous subshock. The
both requirements are crucial for improving high energy particle
confinement and making the shock precursor shorter, in agreement
with the observations. Large scales should be present in the ambient
plasma as a seed for their amplification by the acoustic instability
and could be driven (or seeded) by wave packet modulations. Apart
from that, they result from the coalescence of shocks generated by
the instability, and from the scattering of Alfven waves in
$k$-space by these shocks to larger scales \cite{MD06,DM07,MD09}.
Note that the Bell instability is essentially a short scale
instability (the maximum growth rate is at scales smaller than the
gyro-radii of accelerated particles). At larger scales the magnetic
field growth rate is dominated by the modified resonant and the
longwavelength nonresonant  instabilities \citep[][]{Bykov11}. It
should be noted that vorticity (and thus magnetic field) can be
efficiently generated also at the subshock \citep[see
e.g.][]{mw70,b82,b88,Kevlahan97,KulsrudMF97,GiacJok07,beresnyak09,Fraschetti13}.
This would be too late for improving particle confinement and
reducing the scale of the shock precursor. A more favorable for
acceleration scenario is the above discussed field amplification in
the CR shock precursor.

Now the question is which instability dominates the CR dynamics?
Given the finite precursor crossing time, it is reasonable to choose
the fastest growing mode and consider the development of a slower
one under conditions created by the fast mode after its saturation.
The Bell instability is likely to be efficient at the outskirt of
the shock precursor where the CR current is dominated by the
escaping CRs of the highest energies.  The pressure gradient and the
pitch angle anisotropy are strong enough to drive the acoustic and
resonant instability in the shock precursor \citep[see
e.g.][]{Pelletier06}. Recall that the anisotropy is typically
inversely proportional to the local turbulence level which is
usually decrease with the distance from the shock

Within the main part of the shock precursor, both the CR-pressure
gradient and CR current are strong, so that the nonresonant
CR-driven  instabilities are likely to be the strongest candidates
to govern the shock structure. In fact, these instabilities are
coupled, not only by the common energy source but also dynamically.
But first, it is important to identify conditions under which one of
the instabilities dominates.

\section{Cosmic plasmas with cosmic rays:~the governing equations}

In this section we discuss the governing equations for MHD-type
flows of a cold background plasma interacting with cosmic rays. In
most cases the cosmic ray particles are not subject to binary Coulomb
or nuclear interactions with the background plasma
particles. The interaction between the two components is due to both
regular and fluctuating electromagnetic fields produced by the CRs. The momentum
equation for the background plasma, including the Lorentz force associated with
these fields is given by
\begin{equation}\label{eqMotiontot_0}
 \widetilde{\rho}\left(\frac{\partial\mathbf{\widetilde{u}}}{\partial
 t}+(\mathbf{\widetilde{u}}\nabla)\mathbf{\widetilde{u}}\right)
 =- \nabla \widetilde{p}_{g}+  \frac{1}{c}\mathbf{\widetilde{j}}\times\mathbf{\widetilde{B}}+e\left(\widetilde{n}_{p}-\widetilde{n}_{e}\right)\mathbf{\widetilde{E}},
\end{equation}
where $\mathbf{\widetilde{B}}$ is the magnetic field induction,
$\mathbf{\widetilde{E}}$ - the electric field,
$\mathbf{\widetilde{u}}$ - the bulk plasma velocity,
$\widetilde{p}_{g}$ - the plasma pressure, $\mathbf{\widetilde{j}}$ -
the  electric current carried by the background plasma.
We assume quasi-neutrality for the whole system consisting
of background plasma protons of number density $\widetilde{n}_{p}$, electrons of
number density $\widetilde{n}_{e}$, and cosmic rays of number density $\widetilde{n}_{cr}$.
For simplicity we consider cosmic-ray protons only such that
$\widetilde{n}_{p}+\widetilde{n}_{cr} =\widetilde{n}_{e}$, and typically $\widetilde{n}_{cr} \ll \widetilde{n}_{p}$.

The magnetic field is assumed to be frozen into the background
plasma
\begin{equation}\label{eqVmor}
\displaystyle\mathbf{\widetilde{E}}=-\frac{1}{c}\left[\mathbf{\widetilde{u}}\times\mathbf{\widetilde{B}}\right].
\end{equation}
Both the background electric current $\mathbf{\widetilde{j}}$ and
the electric current of accelerated particles
$\mathbf{\widetilde{j}}^{cr}$ are the sources of magnetic fields in
Maxwell's equations, where the Faraday displacement current was
omitted for the slow MHD-type processes

\begin{equation}\label{Maxw}
\nabla\times\mathbf{\widetilde{B}}
=\displaystyle\frac{4\pi}{c}\left(\mathbf{\widetilde{j}}+\mathbf{\widetilde{j}}^{cr}\right).
\end{equation}
Then, for the quasi-neutral background plasmas, using
Eq.(\ref{eqMotiontot_0}), Eq.(\ref{eqVmor}) and Eq.(\ref{Maxw}),
one can write the induction equation and the equation of motion of
the background plasma in the form used by
\citet[][]{Bell04,Bykov11,Schure11}

\begin{equation}\label{largeInd0}
\frac{\partial\mathbf{\widetilde{B}}}{\partial t}=\nabla
\times(\mathbf{\widetilde{u}}\times \mathbf{\widetilde{B}}),
\end{equation}
\begin{equation}\label{eqMotiontot0}
 \widetilde{\rho}\left(\frac{\partial\mathbf{\widetilde{u}}}{\partial
 t}+(\mathbf{\widetilde{u}}\nabla)\mathbf{\widetilde{u}}\right)
 =- \nabla \widetilde{p}_{g}+  \frac{1}{4\pi}(\nabla\times\mathbf{\widetilde{B}})\times\mathbf{\widetilde{B}} -
\frac{1}{c}(\mathbf{\widetilde{j}}^{cr}-e\widetilde{n}_{cr}\mathbf{\widetilde{u}})\times\mathbf{\widetilde{B}}.
\end{equation}

The microscopic CR-dynamics can be described by a
kinetic equation for the single-particle distribution function
$\widetilde{f}$ that has the form
\begin{equation}\label{KinEqCR}
\frac{\partial\widetilde{f}}{\partial
t}+\mathbf{v}\cdot\frac{\partial \widetilde{f}}{\partial\mathbf{r}}+
e\mathbf{\widetilde{E}}\cdot\frac{\partial\widetilde{f}}{\partial\mathbf{p}}-
\frac{ec}{\mathcal{E}}\mathbf{\widetilde{B}}
\cdot\widehat{\mathbf{\mathcal{O}}}\widetilde{f}=0,
\end{equation}
where the CR particle energy is $\mathcal{E}$,
$\widehat{\mathbf{\mathcal{O}}}$ is the momentum rotation operator
\citep[see e.g.][]{Toptygin83,ber12}. There are no Coulomb
collisions in the kinetic equation Eq.(\ref{KinEqCR}), but the
microscopic electromagnetic fields are fluctuating in a wide
dynamical range due to collective plasma effects. The coarse grained
distribution function of the CR particles $f = <\widetilde{f}>$
obeys the equation that can be obtained by averaging the microscopic
equation Eq.(\ref{KinEqCR}) over an ensemble of appropriate
short-scale fluctuations
\begin{equation}\label{KinEqCR1}
\frac{\partial f}{\partial t}+\mathbf{v}\cdot\frac{\partial
f}{\partial\mathbf{r}}+e\mathbf{E}\cdot\frac{\partial
f}{\partial\mathbf{p}}-\frac{ec}{\mathcal{E}}\mathbf{B}
\cdot\widehat{\mathbf{\mathcal{O}}}f=I[f,f^{'}].
\end{equation}
Here $\widetilde{f} = f + f^{'}$,
$\mathbf{\widetilde{B}}=\mathbf{B}+\mathbf{b^{'}}$,
$\mathbf{\widetilde{E}}=\mathbf{E}+\mathbf{E^{'}}$, $\mathbf{B} =
<\mathbf{\widetilde{B}}>$, $\mathbf{E} = <\mathbf{\widetilde{E}}>$ -
are the averaged fields,  and therefore $\left\langle
\mathbf{b^{'}}\right\rangle=0$, $\left\langle
\mathbf{E^{'}}\right\rangle=0$. The ensemble of fluctuations can be
of external origin or produced by the same population of charged
particles we only assumed at this point that the collision operator
\begin{equation}\label{integrSt}
I[f,f^{'}] = -e\left\langle\mathbf{E^{'}}\cdot\frac{\partial
f^{'}}{\partial\mathbf{p}}\right\rangle+
\frac{ec}{\mathcal{E}}\left\langle\mathbf{b^{'}}\cdot\widehat{\mathbf{\mathcal{O}}}f^{'}\right\rangle,
\end{equation}
is a functional of the averaged distribution function $f$ and can be
expressed through the statistical momenta of the fluctuating field.
The collision operator describes the momentum and energy exchange
between CRs and the background plasma and therefore it must be
accounted for in the averaged governing equations for both the CRs
and background plasma.

The momentum exchange rate is the first moment of
Eq.(\ref{integrSt})
\begin{equation}\label{integrSt1}
\int \mathbf{p}I[f]d^{3}p=-e\left\langle
n_{cr}^{'}\mathbf{E^{'}}\right\rangle+\frac{1}{c}\left\langle\mathbf{j}_{cr}^{'}\times
\mathbf{b^{'}}\right\rangle,
\end{equation}
where $n_{cr}^{'}$, $\mathbf{j}_{cr}^{'}$ - are the fluctuating
parts of the CR number density and the CRs electric current defined
by
\begin{equation}\label{densSt1}
n_{cr}^{'}=e\int f^{'} d^{3}p,
\end{equation}
\begin{equation}\label{currentSt1}
\mathbf{j}_{cr}^{'}=e\int\mathbf{v}(p) f^{'} d^{3}p,
\end{equation}
where $\mathbf{v}(p)$ - is the CR particle velocity, and
$\left\langle f^{'}\right\rangle=0$.

Then, by averaging the last term in Eq.(\ref{eqMotiontot0}), one can
get
\begin{equation}\label{intStFon}
\frac{1}{c}\left\langle\left(\mathbf{\widetilde{j}}_{cr}-e\nCR\mathbf{\widetilde{u}}\right)\times
\mathbf{\widetilde{B}}\right\rangle=\frac{1}{c}(\mathbf{j}^{cr}-en_{cr}\mathbf{u})\times
\mathbf{B}-e\left\langle
n_{cr}^{'}\mathbf{E^{'}}\right\rangle+\frac{1}{c}\left\langle\mathbf{j}_{\rm
cr}^{'}\times \mathbf{b^{'}}\right\rangle,
\end{equation}
where $n_{\rm cr}$, $\mathbf{j}_{\rm cr}$ - are the averaged CR
number density  and their electric current,
$\mathbf{\widetilde{j}}_{cr}=\mathbf{j}_{\rm cr}+\mathbf{j}_{\rm
cr}^{'}$, $\widetilde{n}_{cr}=n_{\rm cr}+n_{\rm cr}^{'}$. Note that
Eq.(\ref{integrSt1}) and the last two terms on the right hand side of
Eq.(\ref{intStFon}) are coincident. Therefore, we conclude that the
CR scattering due to the stochastic electromagnetic fields accounted
for in the kinetic equation Eq.(\ref{KinEqCR1}) by the collision
operator must be simultaneously included into the equation of motion
of the background plasma using Eq.(\ref{intStFon}).

The averaged induction equation Eq.(\ref{largeInd0}) can be
expressed as
\begin{equation}\label{largeInd}
\frac{\partial\mathbf{B}}{\partial t}=\nabla \times(\mathbf{u}\times
\mathbf{B}),
\end{equation}
and the averaged equation of motion Eq.(\ref{eqMotiontot0}) for the
background plasma
\begin{equation}\label{eqMotiontot}
\rho\left(\frac{\partial\mathbf{u}}{\partial
 t}+(\mathbf{u}\nabla)\mathbf{u}\right)
 =- \nabla p_{g}+  \frac{1}{4\pi}(\nabla\times\mathbf{B})\times\mathbf{B} -
\frac{1}{c}(\mathbf{j}^{cr}-en_{\rm
cr}\mathbf{u})\times\mathbf{B}-\int \mathbf{p}I[f]d^{3}p,
\end{equation}
where $p_{g}$ -is the averaged pressure of background plasma. Note
that Eq.(\ref{largeInd}) and Eq.(\ref{eqMotiontot}) is also valid
for CRs  consisting of electrons and positrons, with
$n_{\rm cr}$ being the difference between the positron and the
electron number densities, while  $\mathbf{j}^{cr}$ -- the total
electric current of the particles.

In a few cases, namely, for weakly  fluctuating
magnetic fields or, for strong magnetic fluctuations but at
scales smaller than the CR gyroradii,  some closure procedures exist
to reduce  the collision operator $I[f,f^{'}]$  to $I[f]$ \citep[see
e.g.][]{Toptygin83,ber12}. It is instructive, nevertheless, to
derive the force density $\int \mathbf{p}I[f]d^{3}p$ for the most
simple case of  $I[f]$. The simplest form of  the collision operator
is the relaxation time approximation in the rest frame of the
background plasma
\begin{equation}\label{integrSt2}
I[f]=-\nu \left(f- f_{\rm iso}\right),
\end{equation}
where $f_{\rm iso}$ - is the isotropic  part of the  momentum
distribution $f$, and $\nu$ is the collision frequency due to CR
particle-wave interactions \citep[e.g.][]{Bykov11}. This approach
usually implies that the scatterers have no mean (or drift) velocity
relative to the rest frame of the background plasma. This is not
always true, if the plasma instabilities that are producing the
magnetic field fluctuations are highly anisotropic. However, it can
be used to illustrate the importance of  the momentum exchange
between CRs and the background plasma.

Using the parameterisation  $\nu=a\Omega$, where
$\Omega=\displaystyle\frac{ecB_{0}}{\mathcal{E}}$, $\mathbf{B}_{0}$
is the mean magnetic field, and $a$ - is the CR collisionality
parameter, from Eq.(\ref{integrSt2}), one can obtain
\begin{equation}\label{integrSt3}
\int \mathbf{p}I[f]d^{3}p=-\frac{aB_{0}}{c}\mathbf{j}_{\rm cr}.
\end{equation}
This is the force density in Eq.(\ref{eqMotiontot}).

\section{Instabilities driven by anisotropic CR distributions:\\ the kinetic approach}\label{kin_appr}
Consider incompressible modes propagating along the mean homogeneous
magnetic field $\mathbf{B}_{0}$ in the rest frame of the background
plasma. The linear dispersion relation can be obtained by the
standard perturbation analysis of Eq.(\ref{largeInd}),
Eq.(\ref{eqMotiontot}) and Eq.(\ref{KinEqCR1}), assuming the small
perturbations of magnetic field $\mathbf{b}$, plasma bulk velocity
$\mathbf{u}$ and the CR distribution  $f$ to be
$\propto\exp\left(ikx- i\omega t\right)$. The unperturbed
anisotropic CR distribution, that is the source of the instability
free energy, can be represented as
\begin{equation}\label{distrF0}
f^{cr}_{0}=\frac{n_{cr}N\left(p\right)}{4\pi}\left[1+3\beta\mu+\frac{\chi}{2}\left(3\mu^{2}-1\right)\right],
\end{equation}
where  $\mu=\cos\theta$ , $\theta$ - is the CR particle pitch-angle,
 $n_{\rm
cr}$- CR number density. The multipole moments of the CR angular
distribution are parameterized by $\beta$ (the dipole) and $\chi$
(the quadrupole). We assume below  $\beta \leq$ 1  and $\chi \leq$
1. The unperturbed state can be a steady state of a system with CRs
where both the anisotropy and the spectral distribution $N(p)$ are
determined by  the energy source and sink as well as the magnetic
field geometry through the kinetic equation Eq.(\ref{KinEqCR1}) with
some appropriate boundary conditions. The most interesting
application of the formalism is related to diffusive shock
acceleration model \citep[see
e.g.][]{Blandford87,MDru01,ber12,schureea12}. In that case the
normalized power-law CR spectrum is appropriate:
\begin{equation}\label{spektrNp}
N\left(p\right)=\frac{\left(\alpha-3\right)p_{0}^{\left(\alpha-3\right)}}{\left[1-\left(\frac{p_{0}}{p_{m}}\right)^{\alpha-3}\right]p^{\alpha}},\,\,
p_{0}\leq p\leq p_{m},
\end{equation}
where  $\alpha$ - is the spectral index,  $p_{0}$ and $p_{m}$  - are
the minimal and maximal CR momenta, respectively. In the DSA
applications it is convenient to express the dipole anisotropy
parameters through the shock velocity $u_{\rm s}$ as $\displaystyle
\beta = \frac{u_{\rm s}}{c}$.

Then dispersion equation has the form:
\begin{equation}\label{dispers}
\omega^{2}=v_{a}^{2}\left\{k^{2}\mp k\left[\left(1\pm
ia\right)\left(k_{0}A_{0}\left(x_{0},x_{m}\right)+\frac{4\pi
en_{cr}\chi}{B_{0}}A_{1}\left(x_{0},x_{m}\right)\right)-k_{0}\right]\right\},
\end{equation}
where $v_{a}=\displaystyle\frac{B_{0}}{\sqrt{4\pi\rho}}$,
$k_{0}=\displaystyle\frac{4\pi}{c}\frac{j_{0}^{cr}}{B_{0}}$,
$j_{0}^{cr}=en_{cr}u_{\rm s}$, $\displaystyle  x=\frac{kcp}{eB_{0}}$
, $\displaystyle x_{0}=\frac{kcp_{0}}{eB_{0}}$ , $\displaystyle
x_{m}=\frac{kcp_{m}}{eB_{0}}$,
\begin{equation}\label{koeffA}
A_{0,1}\left(x_{0},x_{m}\right)=\int_{p_{0}}^{p_{m}}\sigma_{0,1}\left(p\right)N\left(p\right)p^{2}dp
\end{equation}
\begin{equation}\label{sigma0Int}
\sigma_{0}\left(p\right)=\frac{3}{4}\int_{-1}^{1}\frac{\left(1-\mu^{2}\right)}{1\mp
x\mu\pm ia}d\mu,
\end{equation}
\begin{equation}\label{sigma1Int}
\sigma_{1}\left(p\right)=\frac{3}{4}\int_{-1}^{1}\frac{\left(1-\mu^{2}\right)\mu}{1\mp
x\mu\pm ia}d\mu,
\end{equation}
where the  $\pm$ signs correspond to the two possible circular
polarizations defined by $\mathbf{b}= b(\mathbf{e}_{y}\pm
i\mathbf{e}_{z})$, with the $x$-axis along the mean field
$\mathbf{B}_{0}$. The functions $A_{0,1}\left(x_{0},x_{m}\right)$
are expressed in elementary functions in Appendix A. In the
collisionless limit $a \rightarrow 0$  the contribution of the pole
to the imaginary part of Eq.(\ref{sigma0Int}) describes the well
known resonant instability \citep[e.g.][]{zweibel79,amato11}, while
the real part (the principal part of the integral) is responsible
for the instability discovered by \citet{Bell04} (see also
\citet{achterberg83}).

The kinetic approach we used here to derive the dispersion equation
 allows us to unify the instabilities due to both the dipole and
quadrupole-type CR  anisotropy. The finite mean free path of the CRs
is characterized by the collisionality parameter $a$. The approach
used above allows one to study the instabilities driven by the CR
anisotropy for arbitrary relations between the mode wavelength, the
CR mean free path and the CR gyroradii. It is instructive to
demonstrate the transition between the collisionless case (i.e.
$a=0$),  where the CR mean free part is much larger than the mode
wavelength, and the opposite case  with the collisionality parameter
$a\rightarrow1$ (Bohm's diffusion limit).   In the collisionless
limit (i.e. $a=0$) the instabilities due to dipole type anisotropy
($\chi =0$) were discussed by \citet{Bell04}, \citet{Pelletier06},
and \cite{Amato09}. The firehose instability of a highly
relativistic plasma without a dipole anisotropy was discussed by
\citet{Noerdlinger68}. \citet{Schure11} derived a dispersion
equation for the mono-energetic particle distribution instead of the
power-law distribution in Eq.(\ref{spektrNp}) used here, and the
dipole-type initial anisotropy (i.e. $\chi$ =0).  The firehose
instability of the anisotropic CR pressure with nonzero $\chi $  was
studied by \citet{Bykov11a}.

\section{Growth rates of incompressible modes propagating along the
mean magnetic field} 
\begin{figure*}
\centering
\includegraphics[width=10cm]{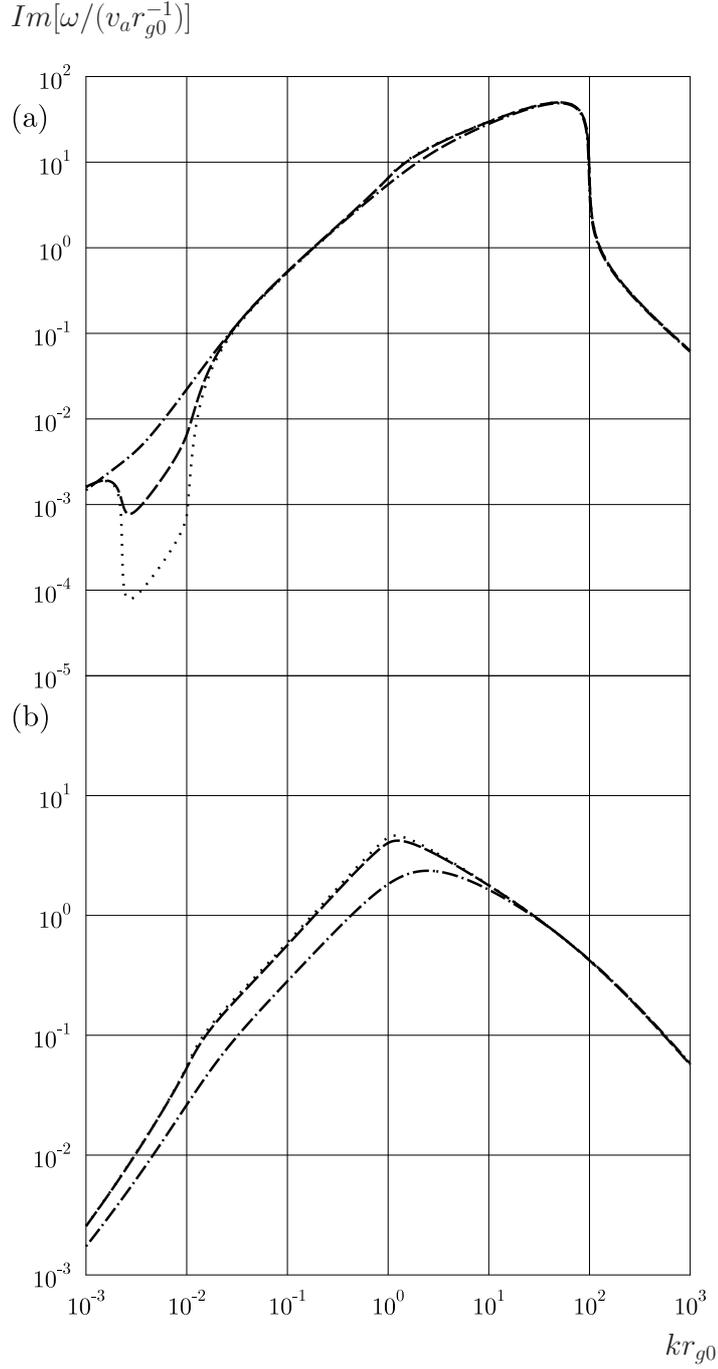}
\caption{The growth rates for the two circularly polarized modes.
The right hand polarized mode (panel $a$) and the left hand mode
(panel $b$) are derived from Eq.(\ref{dispers}). We illustrate the
growth rate dependence on the collisionality parameter $a$. Dotted
line corresponds to $a$=0.01, dashed line - $a$=0.1, and dot-dashed
line -  $a$=1. The quadrupole anisotropy is $\chi=6\,(u_{\rm
s}/c)^2$. Note that in the bottom panel the dashed and dotted lines
are very close.} \label{figSh}
\end{figure*}

In Figure \ref{figSh} we illustrate the growth
rates derived from Eq.(\ref{dispers}) for a particular choice of
parameters of the CR distribution functions typical for the upstream
distribution of CRs accelerated by the diffusive acceleration at a
shock of velocity $\displaystyle\frac{u_{\rm s}}{c}=0.01$, with
$\alpha=4$, and $\displaystyle\frac{p_{m}}{p_{0}}=100$. The DSA spectrum
 may span many decades, but we choose the two-decade range of the particle spectrum to
model the instability far upstream of the shock where the
longwavelength fluctuation amplification is the most efficient. The
CR distribution function and the CR current normalizations are fixed
here by the dimensionless parameter $k_{0}r_{g0}=100$, where
$\displaystyle r_{g0}=\frac{cp_{0}}{eB_{0}}$.To estimate the
normalization of the CR distribution we assumed that about 10\% of
the shock ram pressure is converted into the CR energy. For the CR
spectrum of the index $\alpha=4$ the fraction of CRs above the
momentum $p_{0}$ is $\propto p_{m}/p_{0}$, while $r_{g0} \propto
p_{0}$. Therefore the spatial dependence of the key governing
parameter of the Bell instability  $k_{0}r_{g0}$ depends basically
on the energy dependent CR anisotropy. The bulk of the CRs  are
confined in the accelerator and therefore would have anisotropy
about $u_{\rm s}/c$ (apart from the particles at the very end of the
CR spectrum escaping from the system).

\subsection{Nonresonant shortwavelength instability}
It is instructive to consider  the short-scale CR-current driven
modes produced by Bell's  instability as an asymptotic case of the
general Eq.(\ref{dispers}), for different wavenumbers $k$ in the
collisionless case $a$ = 0, following \citet{Bell04} and
\citet{Bykov11a}. In Figure \ref{figSh}, we illustrate the growth
rate dependence on the collisionality parameter.

In the wavenumber range $k_{0}r_{g0} >kr_{g0}>1$, corresponding to
the  instability discovered by \citet{Bell04}, the growth of the
right hand polarized mode (panel \rm{a} in Figure \ref{figSh}) is
much faster than the left hand mode (panel {\it b} in Figure
\ref{figSh}). This results in fast helicity production. In the
collisionless limit the right hand mode has the growth rate
\begin{equation}\label{gam_Bell}
\gamma_{b}=v_{a}\sqrt{k_{0}k-k^{2}}.
\end{equation}
Eq.(\ref{gam_Bell}) follows from Eq.(\ref{dispers}), neglecting the
response of the CR current on the magnetic fluctuations, i.e.,
$A_{0}(x_{0},x_{m}) \rightarrow 0$ and $A_{1}(x_{0},x_{m})
\rightarrow 0$. The weak CR-current response is the main cause of
the Bell-type instability. Indeed, the CR current induces the
compensatory reverse current in the background plasma and if the
current is not responding to a magnetic field variation, then the
magnetic fluctuation is growing due to the Ampere force. The CR
current only weakly responds to the magnetic field fluctuations with
wavenumbers $k_{0}r_{g0}
>kr_{g0}> 1$, and they grow. From
Eq.(\ref{gam_Bell}) one may see that $\displaystyle\gamma_{b}\sim
k^{\frac{1}{2}}$ for $k\ll k_{0}$.

\subsection{The resonant instability}
In the collisionless case for the wavenumber regimes $x_{m}>1$, but
$x_{0}<1$,  the resonant contribution dominates the pole in the
integrand in Eq.(\ref{sigma0Int}). Therefore, the resonant mode
growth can be seen in Figure \ref{figSh} in the regime
$0.01<kr_{g0}<1$, where both circular polarization modes are growing
with the very close rates  $\propto k$ for $\alpha=4$ (compare
panels {\it (a)} and {\it (b)} in Figure \ref{figSh}). Collisions do
not change the mode growth drastically for $a<0.1$, but in the limit
of strong collisions with $a=1$ the left hand mode grows slower than
the right hand polarized mode. This may also result in helicity
production.

\subsection{A nonresonant longwavelength instability: the firehose mode}
In the longwavelength regime where $\displaystyle
x_{m}=\frac{kcp_{m}}{eB_{0}}\ll 1$, within the collisionless case,
the dispersion relation in Eq.(\ref{dispers}) can be approximated,
following \citet{Bykov11a}, as
\begin{equation}\label{dispers2asimpX0}
\omega^{2}=v_{a}^{2}k^{2}\left\{1\mp
\frac{r_{g0}}{5}\left[k_{0}x_{m}\pm\frac{4\pi
en_{cr}\chi}{B_{0}}\frac{\ln\frac{p_{m}}{p_{0}}}{\left(1-\frac{p_{0}}{p_{m}}\right)}\right]\right\}.
\end{equation}
As it follows from Eq.(\ref{dispers2asimpX0}), in the regime
dominated by the dipole CR anisotropy ($\chi \rightarrow 0$) only
the left-polarized mode is growing with the rate $\propto
\displaystyle k^{\frac{3}{2}}$ \citep[see][]{Schure11}. For a finite
quadrupole-type CR anisotropy $\chi$  at small enough wavenumbers
 the modes of both circular polarizations
are growing again with the very close rates $\propto k$ (see in
Figure \ref{figSh}). The instability due to the quadrupole-type CR
anisotropy corresponds to the well known firehose instability in a
plasma with anisotropic pressure. Indeed, the CR pressure anisotropy
derived from the CR distribution Eq.(\ref{distrF0}) is
\begin{equation} \label{anisotrPcr}
P_{\parallel}^{cr}-P_{\perp}^{cr}=\frac{3}{5}\chi P^{cr},
\end{equation}
where
\begin{equation}\label{Pcr}
P^{cr}=\frac{1}{3}n_{cr}\int_{0}^{\infty}v\left(p\right)N\left(p\right)p^{3}dp.
\end{equation}
The dispersion relation for the modes produced by only the
quadrupole-type anisotropy of CR distribution can be obtained from
Eq.(\ref{dispers2asimpX0}) if one neglects the dipole-type
contribution $x_{m}\rightarrow 0$. Then, it is reduced to the
standard hydrodynamic dispersion relation of the firehose
instability
\begin{equation}\label{dispersMHD}
\omega=\pm\sqrt{v_{a}^{2}-\frac{P_{\parallel}-P_{\perp}}{\rho}}k,
\end{equation}
where $P_{\parallel}-P_{\perp}$  - is the pressure anisotropy along
the mean magnetic field direction \citep[see,
e.g.,][]{Blandford87,tb97}. The dispersion relation
Eq.(\ref{dispersMHD}) is justified for the modes with the
wavenumbers above the CR ion gyroradii.  The dependence of the
growth rates of the firehouse instability on the collisionality
parameter can be seen in Figure \ref{figSh}. It should be noted that
the growth rates of the firehose modes of both polarizations in the
regime $kr_{g0}<1$ are declining functions of the collisionality
parameter. Their growth rates would be equal in the case of lack of
the CR current. Contrary, the growth rates of the current driven
modes are different for the two polarizations. The growth rate of
the right hand polarized CR-current driven mode is sensitive to the
collisionality parameter \citep[see][]{Schure11}.

\begin{figure*}
\centering
\includegraphics[width=10cm]{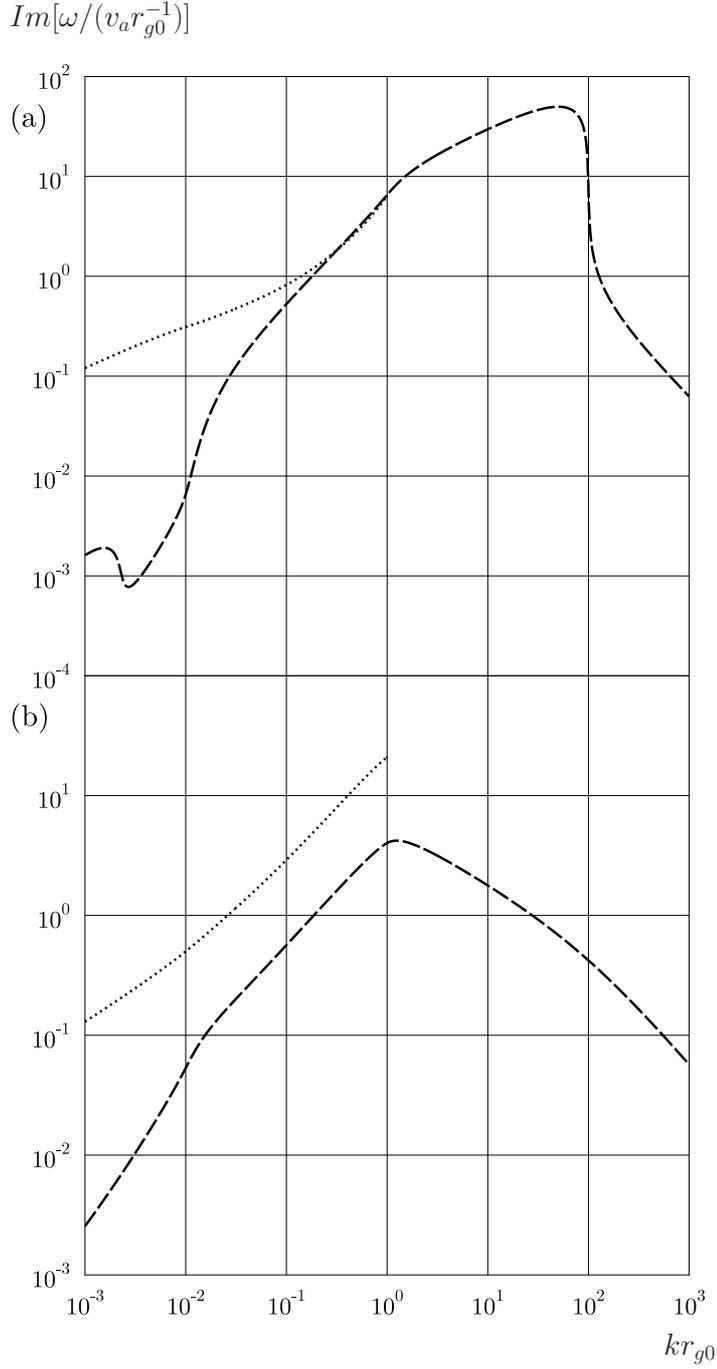}
\caption{The growth rates of the longwavelength modes of two
circular polarizations. The right hand polarized mode (panel $a$)
and the left hand mode (panel $b$) are propagating along the mean
magnetic field as function of the wavenumber. The dotted line curves
are derived from the dispersion equation Eq.(\ref{dispersMedium})
for the collisionality parameter $a$ = 0.1, the dimensionless {\it
r.m.s.} amplitude of Bell's turbulence $N_{B}=1$, and the mixing
parameter $\xi=3$. The dashed curves given for comparison are the
growth rates derived from Eq.(\ref{dispers}) which are shown in
Figure \ref{figSh}.} \label{figNash}
\end{figure*}

\subsection{A nonresonant long-wavelength instability: the cosmic-ray current driven
dynamo}\label{dynamo}
Bell's instability results in the fast growth of short-scale modes
with wavelengths shorter than the gyroradius of the cosmic-ray
particles and in the presence of CR-current it may produce strong
short-scale turbulence
\citep[e.g.][]{BelLuc01,Bell04,zp08,zpv08,rsdk08,vbe09,rogachevskiiea12}.
Moreover, the shortscale turbulence is helical, and at the
wavenumbers below $2\,k_{0}$ its kinetic energy density dominates
over the magnetic energy density making a favorable condition for a
pure $\alpha$-dynamo effect \citep[see][]{Bykov11}. The strong
short-scale turbulence influences the background plasma dynamics on
scales larger than the CR gyroradii. \citet{Bykov11} derived the
mean field dynamic equations averaged over the ensemble of
short-scale motions for plasma systems with CR-current. The averaged
equation of motion can be presented as
\begin{eqnarray} \label{eqMotionNash}
& & \frac{ \partial\mathbf{V}}{\partial t } +
(\mathbf{V}\nabla)\mathbf{V} = -\frac{1}{\rho}\nabla P_g
-\left\langle(\mathbf{u}\nabla)\mathbf{u}\right\rangle
   +\frac{1}{4\pi\rho}\left\langle(\nabla\times\mathbf{b})\times\mathbf{b}\right\rangle+
 \nonumber \\
& &
  +\frac{1}{4\pi\rho}((\nabla\times\mathbf{B})\times\mathbf{B})-
  -\frac{1}{c\,\rho}( (\jCRmeanVec -e\, \nCR
\mathbf{V})\times\mathbf{B})-\int \mathbf{p}I[f]d^{3}p,
 \label{largeEqMotion}
\end{eqnarray}
where $\mathbf{V}$ is the mean velocity of the plasma. The magnetic
induction equation for the mean magnetic field $\mathbf{B}$ reads
\begin{equation}\label{largeIndNash}
\frac{\partial\mathbf{B}}{\partial t}=c\nabla\times
\meanEMF+\nabla \times(\mathbf{V}\times
\mathbf{B})+\nu_{m}\triangle\mathbf{B}.
\end{equation}
Here $\meanEMF=\left\langle \mathbf{u}\times
\mathbf{b}\right\rangle$ is the average turbulent  electromotive
force and  $\nu_{m}$ is the magnetic diffusivity.
The averaged
equations Eq.(\ref{eqMotionNash}) and Eq.(\ref{largeIndNash}) are
designed to be applied to the dynamics of modes with scales larger
than $r_{g0}$, i.e., CR particles are magnetized on these scales.

The ponderomotive forces
$\left\langle(\mathbf{u}\nabla)\mathbf{u}\right\rangle$
   and
   $\frac{1}{4\pi\rho}\left\langle(\nabla\times\mathbf{b})\times\mathbf{b}\right\rangle$
in Eq.(\ref{eqMotionNash}) describe the momentum exchange of the
background plasma with the Bell mode turbulence. The averaged
turbulent electromotive force $\meanEMF$ results in the magnetic
induction evolution. It is  important that in the case under
consideration the ponderomotive forces in Eq.(\ref{eqMotionNash})
depend on the CR current through the Bell mode turbulence moments.
To express the electromotive and ponderomotive forces through the CR
current \citet{Bykov11} followed the mean field closure procedure
similar to the approach proposed by \citet[][]{bf02} in the dynamo
theory \citep[see for a review][]{brandenburg09}. The closure
procedure is introduced by the parameter $\taucor$. The correlation
time $\taucor$ which is the relaxation time of triple correlations
and is approximately equal to the turnover time of the Bell
turbulence. The dependence of the electromotive force and the
ponderomotive force on the CR current is determined by the kinetic
coefficients $\alpt$ and $\kappa_{t}$, correspondingly. The kinetic
coefficients are determined by the r.m.s. amplitude of Bell's
turbulence $\bSqMean$ and $\taucor$. The short scale turbulence
produced by the Bell mode instability is helical and therefore there
is also a contribution to the electromotive force $\propto \alpt
\overline{B}$ resulting in the $\alpha$-dynamo effect. Then, the
dispersion equation for the modes of wavelengths longer than
$r_{g0}$ in a plasma with anisotropic relativistic CRs  can be
derived from Eq.(\ref{eqMotionNash}) and Eq.(\ref{largeIndNash}) by
the standard linear perturbation analysis:
\begin{eqnarray} \label{dispersMedium}
\omega^{2}&-&k^{2}v_{a}^{2}\mp\omega
ik\frac{\alpha_{t}}{4\pi\rho}\left[\frac{1}{2}\left(k_{0}A_{0}\left(x_{0},x_{m}\right)+\frac{4\pi
en_{cr}\chi}{B_{0}}A_{1}\left(x_{0},x_{m}\right)\right)+\frac{3}{2}k_{0}\right]
 \nonumber \\
&\pm& kv_{a}^{2}\left(1+\frac{
\kappa_{t}}{B_{0}}\right)\left[\left(k_{0}A_{0}\left(x_{0},x_{m}\right)+\frac{4\pi
en_{cr}\chi}{B_{0}}A_{1}\left(x_{0},x_{m}\right)\right)-k_{0}\right]
 \nonumber \\
&+&iakv_{a}^{2}\left(k_{0}A_{0}\left(x_{0},x_{m}\right)+\frac{4\pi
en_{cr}\chi}{B_{0}}A_{1}\left(x_{0},x_{m}\right)\right)=0.
\end{eqnarray}
The dispersion relation Eq.(\ref{dispersMedium}) was derived for the
systems where the unperturbed CR-current is directed along the
unperturbed magnetic field, and the short scale turbulence consists
of Bell's modes. It is convenient to introduce two dimensionless
parameters $\displaystyle N_{B}=\frac{\sqrt{\bSqMean}}{B_{0}}$ --
Bell's turbulence r.m.s. amplitude, and the dimensionless mixing
length $\xi$, instead of the correlation time $\taucor$. The mixing
length is defined here as
$2\pi\xi/k_{0}=\tau_{cor}\sqrt{\left\langle
v^{2}\right\rangle}\approx \tau_{cor}\sqrt{ \xi\left\langle
b_{B}^{2}\right\rangle/(4\pi\rho)}$. Then
$\displaystyle\alpha_{t}\approx \bSqMean \taucor \approx 8\pi^2
\sqrt{\xi}N_{B}v_{a} k_{0}^{-1}\rho$ and $\displaystyle
\kappa_{t}=\pi N_{B} B_{0}$.

In Figure \ref{figNash} we illustrate the long wavelength mode
growth derived from Eq.(\ref{dispersMedium}) for $\xi = 3$. The
corresponding mixing length is close to the scale of the maximal
growth rate of the short scale Bell's instability. The
$\alpha$-dynamo effect dominates the growth rate of a polarized mode
shown in Figure \ref{figNash} (panel b) in the intermediate
wavenumber regime $a < kr_{g0}< 1$. One should have in mind that in
the case of Bohm's CR diffusion  $a \sim 1$ and therefore the
intermediate wavenumber regime is rather limited. It should be noted
that the helicity of the unstable, long-wavelength mode studied
above is opposite to that of the short-scale Bell mode. This
provides, at least in principle, the possibility of balancing the
global helicity of the system by combining short and long-wavelength
modes. Care must be taken however, since numerical models indicate a
high saturation amplitude of the Bell mode making a nonlinear
analysis necessary to address the helicity balance issue.  We will
discuss some  nonlinear simulations below in \S\ref{sim}.

\subsection{The cosmic-ray current driven instability in the hydrodynamic
regime}\label{LW} The nonresonant  modes in a hydrodynamic regime,
where the wavelength is longer than the mean free path, i.e.,
$kr_{g0} < a$, are unstable, as it follows from
Eq.(\ref{dispersMedium}) \citep[see][for details]{Bykov11}. Both
circular polarizations in panels {\it a} and {\it b} in Figure
\ref{figNash} grow with the same rate given by
\begin{equation}\label{gam1}
\gamma\approx \sqrt{\frac{\pi\BellAmp}{2}}\sqrt{kk_{0}a}v_{a}.
\end{equation}
The transition from the intermediate wavenumber regime $a < kr_{g0}<
1$, dominated by the dynamo effect discussed in \S\ref{dynamo}, where
the mode growth rate can be approximated by
\begin{equation}\label{gam_alpha}
\gamma\approx 4\pi\sqrt{\xi}N_{B}v_{a}k,
\end{equation}
to the hydrodynamical regime with $kr_{g0} < a$ where the growth
rate is $\propto k^{1/2}$ according to Eq.(\ref{gam1}), is clearly
seen in Figure \ref{figNash} (panel {\it b}). Note that for the mode
polarization shown with the dotted line in the panel {\it a} of
Figure \ref{figNash}, no dynamo-type instability occurs, but the
hydrodynamical regime instability is present. This mode grow fast in
the short wavelength regime $kr_{g0} > 1$ due to Bell's instability.

The effect of the short-scale turbulence on the hydrodynamic regime
instability enters Eq.(\ref{dispersMedium}) through the turbulent
coefficient $\kappa_{t}/B_0$. The turbulent ponderomotive force is
large enough in both the intermediate and hydrodynamical regimes,
and the CR current response in the long-wavelength regime can no
longer be neglected. The current cannot be treated as a fixed
external parameter, as is normally done for the short-scale Bell
instability, and therefore the MHD models of the Bell turbulence
that assume a constant CR-current \citep[see
e.g.][]{BelLuc01,zp08,zpv08,rsdk08,vbe09,rogachevskiiea12} cannot be
directly applied to the nonlinear models of the longwavelength
instabilities discussed above. Particle-in-cell simulations  with
very limited dynamical range performed by \citet[][]{rs09,rs09a}
indicate the importance of the CR backreaction effect on the
CR-driven instabilities. Therefore the  nonlinear dynamics of the
long-wave CR-driven turbulence in a wide dynamical range remains to
be investigated. In the next section we illustrate the nonlinear
evolution of the short scale turbulence driven by a fixed CR
current, using high resolution MHD simulations.

\section{Numerical solutions of the Bell--dynamo instability}\label{sim}

\begin{figure}[b]\begin{center}
\includegraphics[height=4.2cm]{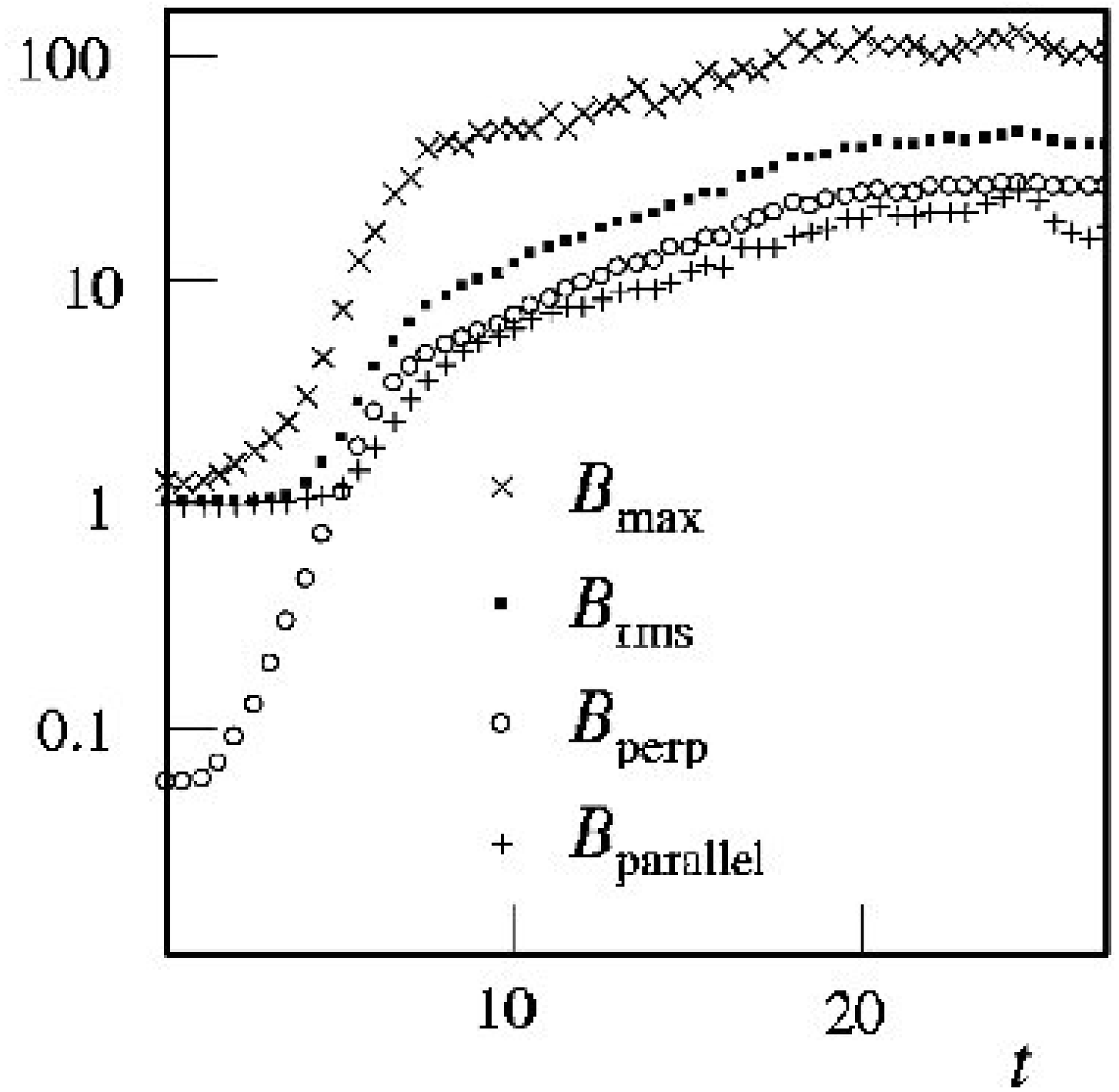}
\includegraphics[height=4.2cm]{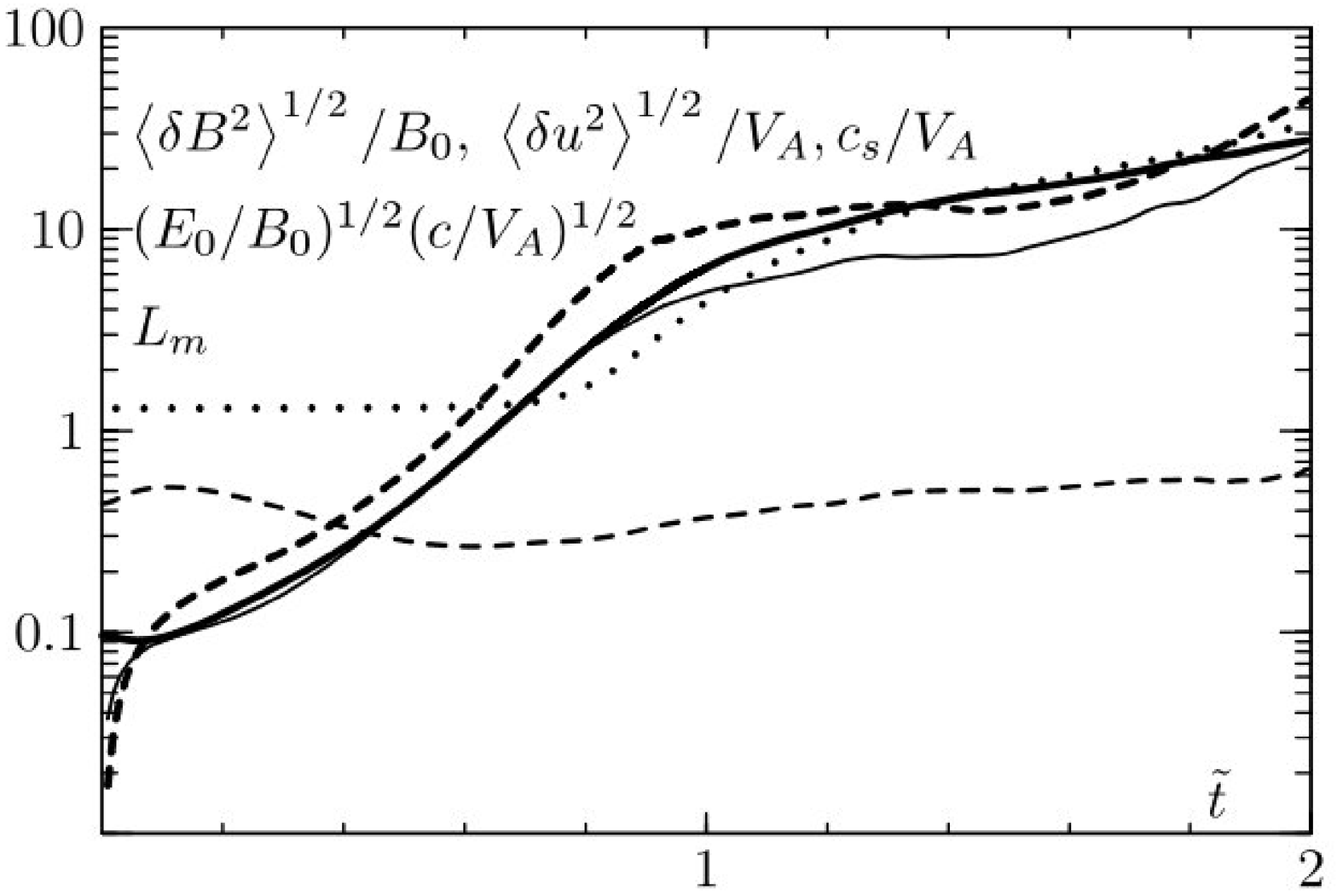}
\end{center}\caption[]{
Numerical solutions of the Bell instability for
${\cal J}=2$ using $128^3$ mesh points \citep[][left hand side]{Bell04} and
${\cal J}=16$ using $256^3$ mesh points \cite[][right hand side]{zpv08}.
Note the continued growth of the magnetic field at the end of the linear
growth phase at $t\approx10$ on the left and $t\approx1$ in the right.
Courtesy of Tony Bell (left panel) and Vladimir Zirakashvili (right panel).
}\label{BellZirakashvili}
\end{figure}

Significant insights have been possible through high-resolution
direct numerical simulations (DNS) and large eddy simulations (LES)
of the Bell instability and its subsequent saturation.
In this section we describe some of the main results and, in particular,
the connection with the dynamo instability.
The simulations have been carried out in a Cartesian domain of
size $L^3$, so the smallest wavenumber in that domain is $k_1=2\pi/L$.
The system is characterized by the non-dimensional parameter
\begin{equation}
{\cal J}={4\pi\over c}{j^{\rm cr}\over k_1 B_0}.
\end{equation}
In the ideal case ($\nu_M=0$), the Bell instability is excited when
${\cal J}>1$ and the normalized wavenumber of the fastest growing mode is
$k/k_1={\cal J}/2$.
The normalized growth rate of this fastest growing mode is
$\gamma_{\it b}/\vAz k_1={\cal J}/2$.
In \Fig{BellZirakashvili} we reproduce the results of numerical
simulations of \cite{Bell04} for ${\cal J}=2$ using $128^3$ mesh points
and \cite{zpv08} for ${\cal J}=16$ using $256^3$ mesh points.
These simulations confirmed the analytically expected linear growth rates.
Interestingly, the saturation of the instability was never perfect.
Instead, the magnetic field still continued to grow at a slow rate.
\cite{rogachevskiiea12} have argued that this slow
growth after the end of the exponential growth phase of the
instability is the result of a mean-field $\alpha$ effect.
The purpose of this section is to elaborate on this possibility.

\begin{figure}
\begin{center}
\includegraphics[width=.49\columnwidth]{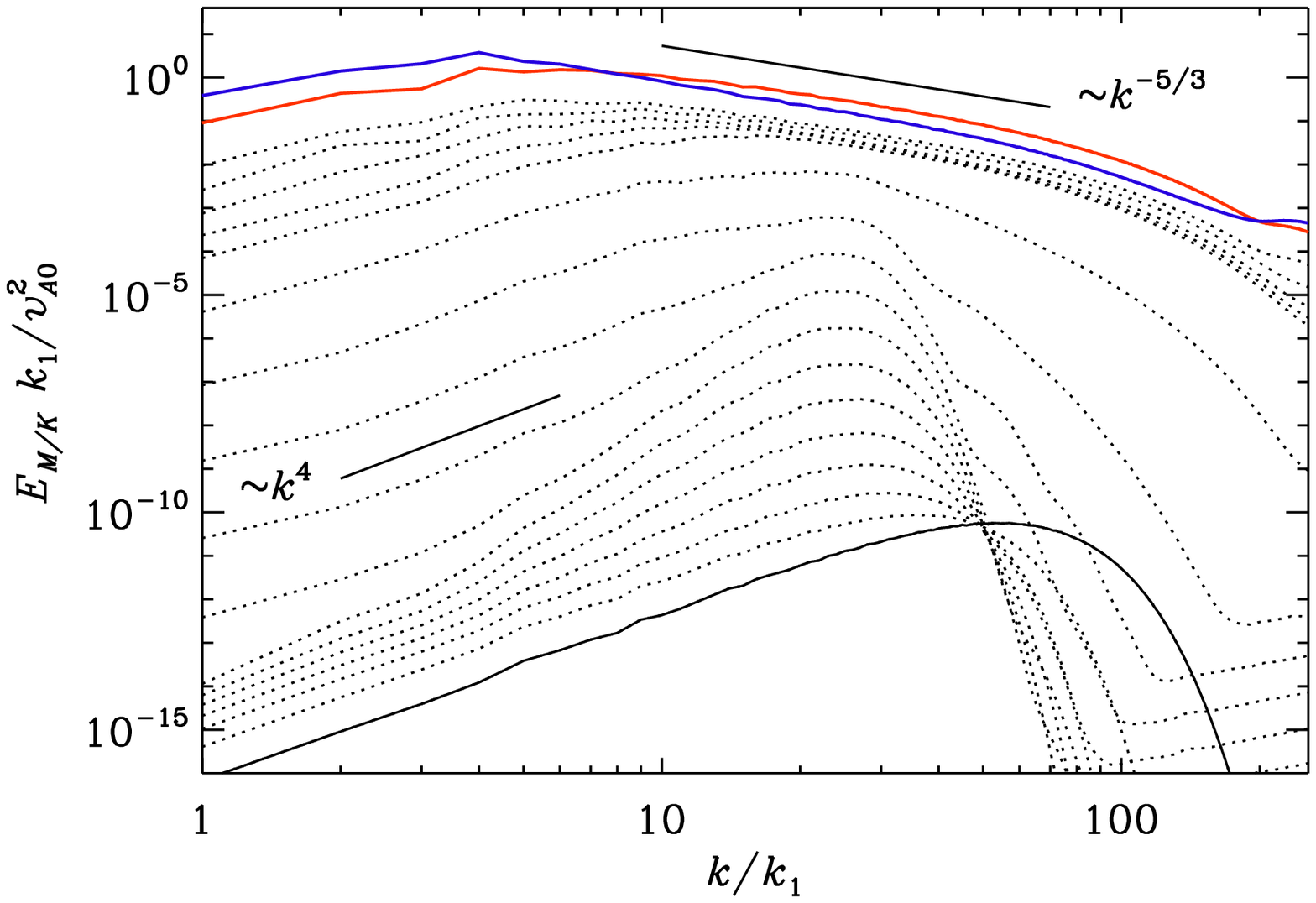}
\includegraphics[width=.49\columnwidth]{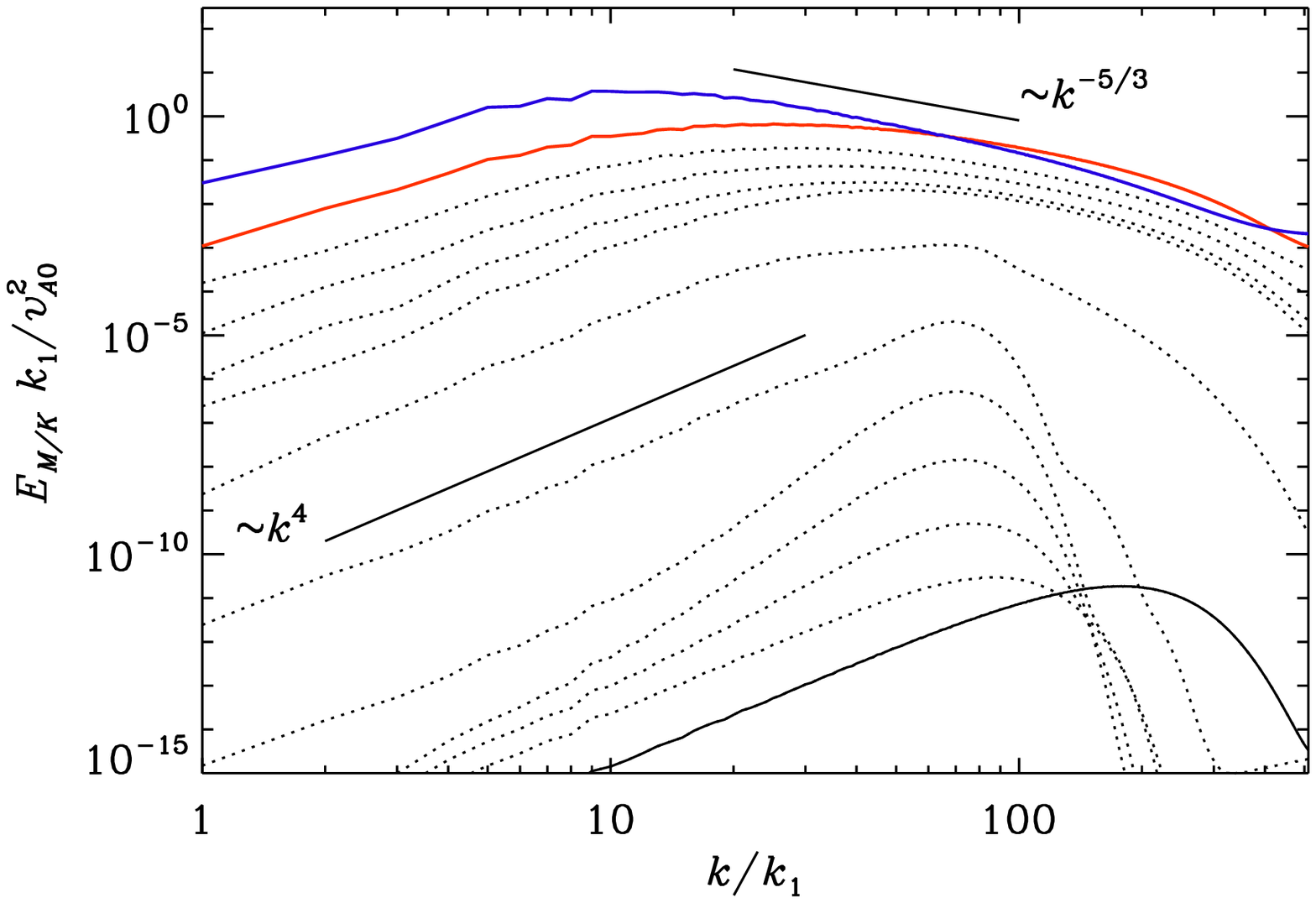}
\end{center}\caption[]{
Time evolution of $E_{\it M}(k,t)$ for ${\cal J}=80$ (left) and
${\cal J}=800$ (right) at resolutions $512^3$ and $1024^3$, respectively.
The solid lines refer to the initial spectra proportional to $k^4$ for
small values of $k$ and the red and blue lines represent the last instant
of $E_{\it M}$ and $E_{\it K}$, respectively.
The straight lines show the $k^4$ and $k^{-5/3}$ power laws.
}\label{ppower_all_EK_EM_T1024eo5axi_powerbz}
\end{figure}

We begin by discussing first the recent DNS of \cite{rogachevskiiea12} for
${\cal J}=80$ and ${\cal J}=800$ at a resolution of $512^3$ mesh points
and discuss also new results for ${\cal J}=800$ at a resolution of
$1024^3$ mesh points.
In all cases, explicit viscosity $\nu$ and magnetic diffusivity
$\nu_{\rm M}$ are used, so the fastest growing modes in those cases have
somewhat smaller wavenumbers than in the ideal case.
This is quantified by the Lundquist number $\Lu=\vA/\nu_{\rm M}k_1$
and the ideal case corresponds then to $\Lu\to\infty$.
For example, \cite{rogachevskiiea12} used $\Lu=80$, in which case
the fastest growing mode has $k_z/k_1\approx21$ for ${\cal J}=80$
while for ${\cal J}=800$ it has $k_z/k_1\approx63$.
The DNS show that most of the power is at somewhat larger wavenumbers;
see \Fig{ppower_all_EK_EM_T1024eo5axi_powerbz}, where we show magnetic
energy spectra for both cases.

\begin{figure}\begin{center}
\includegraphics[width=.49\columnwidth]{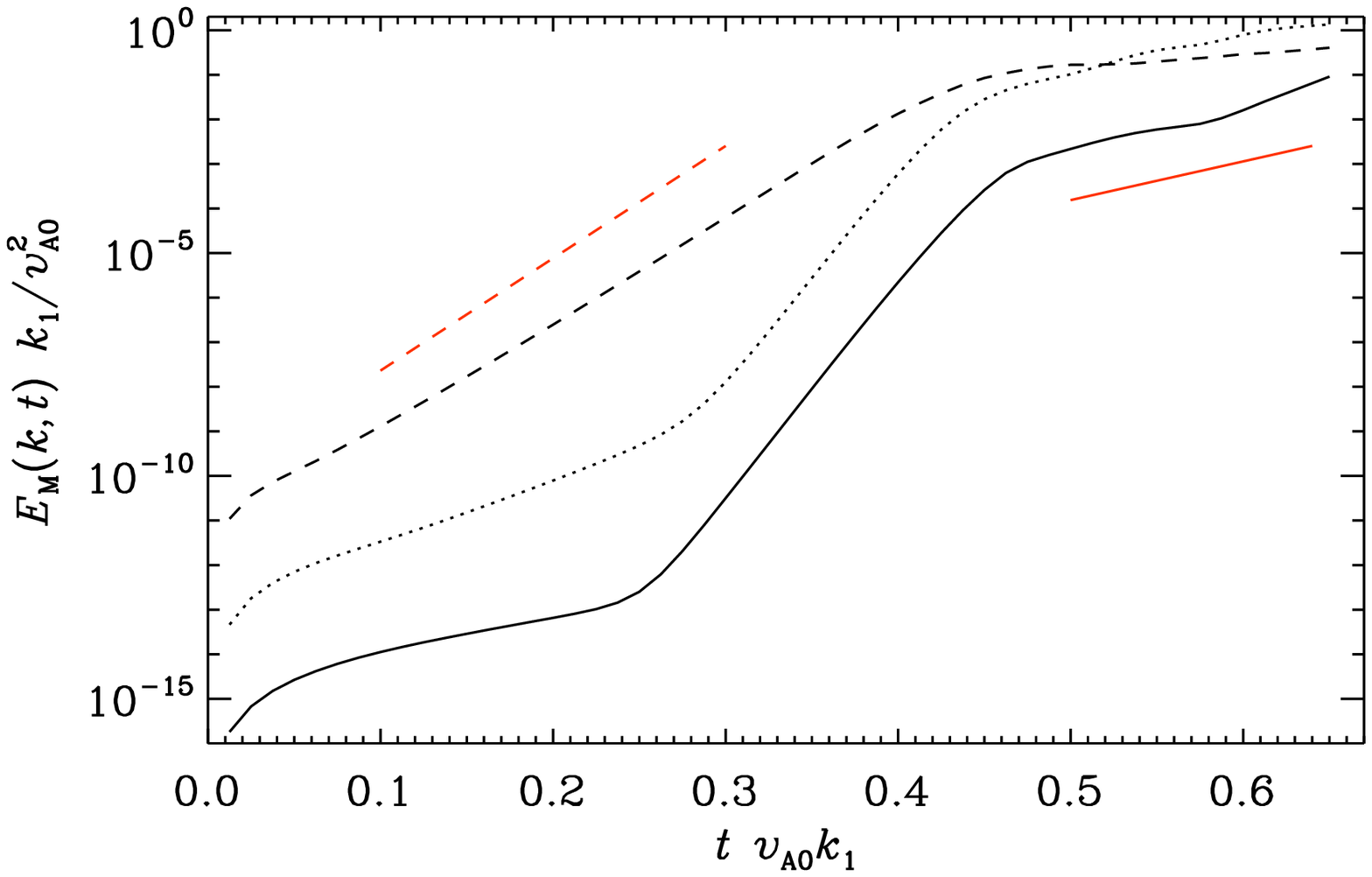}
\includegraphics[width=.49\columnwidth]{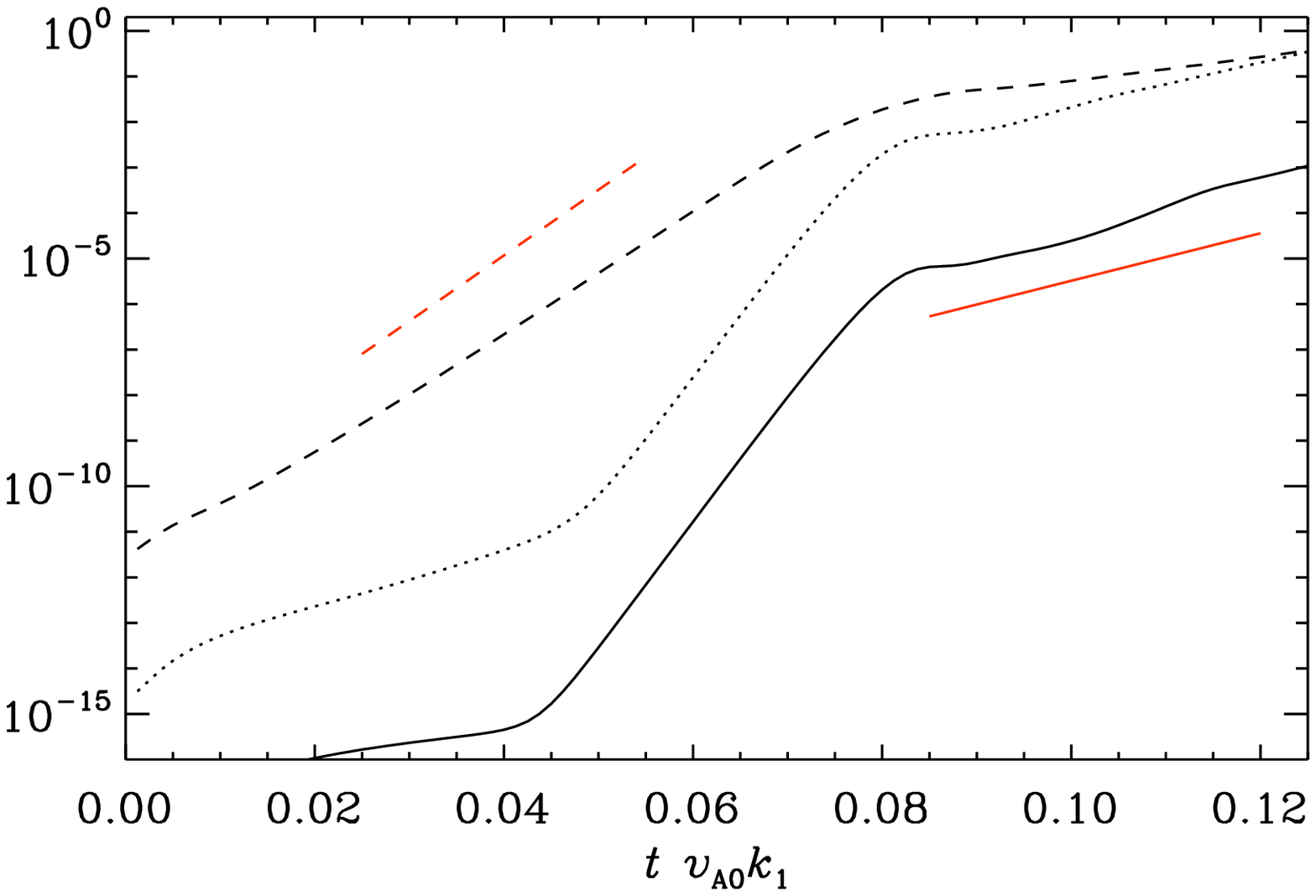}
\end{center}\caption[]{
Time evolution of $E_{\it M} \, k_1/\vAz^2$ for ${\cal J}=80$ (left)
at wavenumbers $k/k_1=1$ (solid line), 5 (dotted), and 21 (dashed)
and ${\cal J}=800$ (right) at wavenumbers $k/k_1=1$ (solid line), 10
(dotted), and 63 (dashed). The short straight lines show the growth
of the energies for the Bell (dashed) and dynamo (solid)
instabilities. }\label{pspec_compt2_tot800}
\end{figure}

In \Fig{pspec_compt2_tot800} we show the temporal evolution of
spectral magnetic energy $E_M$ and the spectral kinetic energy $E_K$
at selected wavenumbers. These curves show an exponential growth at
early times, followed by a slower growth at later times. At the
wavenumbers of the Bell mode, the growth rate from linear theory is
reproduced. At smaller wavenumbers, the growth is at first slower,
and then it is even faster than the growth rate of the Bell mode.
This is a consequence of mode coupling \citep{rogachevskiiea12}.
Comparing with \Fig{ppower_all_EK_EM_T1024eo5axi_powerbz}, we see
that after some time a $k^4$ energy spectrum is established. Such an
energy spectrum is also known as Batchelor spectrum and can be
derived under the constraints of solenoidality and causality
\citep{DC03}. When the $k^4$ spectrum is established, the growth of
spectral energy at small wavenumbers is no longer described by
linear theory, but follows the growth of the Bell mode.

\begin{figure}\begin{center}
\includegraphics[width=\columnwidth]{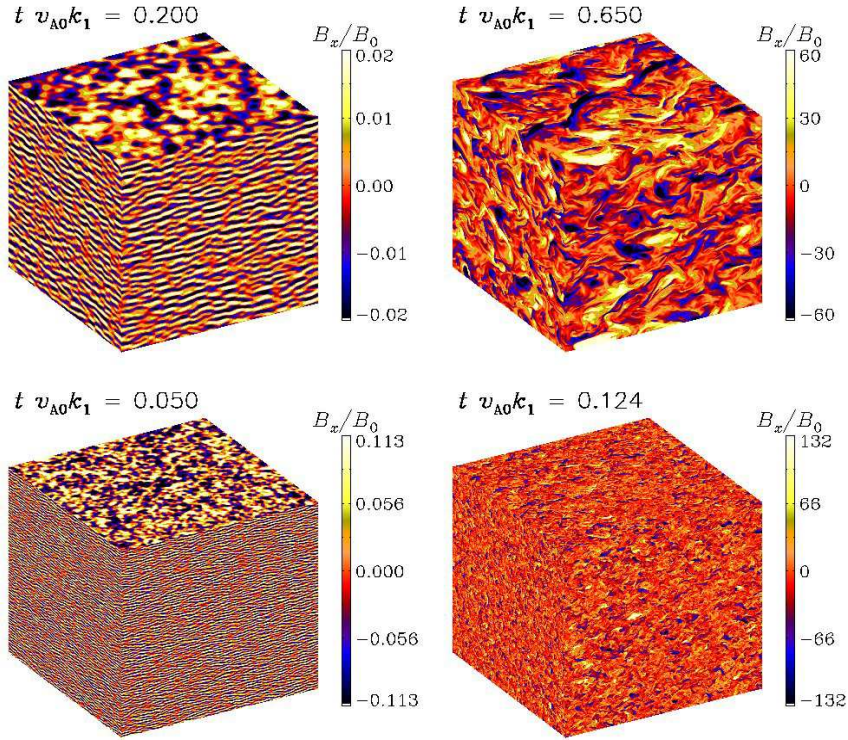}
\end{center}\caption[]{
Visualization of $B_x/B_0$ on the periphery of the computational domain
${\cal J}=80$ using $512^3$ mesh points (upper row) and
${\cal J}=800$ using $1024^3$ mesh points (lower row)
with Lundquist number $\Lu=80$ in both cases.
}\label{Bx_T1024eo5axi}
\end{figure}

In \Fig{Bx_T1024eo5axi} we show visualizations of $B_x/B_0$ on the
periphery of the computational domain for ${\cal J}=80$ using $512^3$
mesh points and ${\cal J}=800$ using $1024^3$ mesh points at two
different times.
One clearly sees that at early times, the magnetic field shows a
layered structure with a high wavenumber in the $z$ direction.
At later times, the magnetic field breaks up and becomes turbulent.
In both cases, larger scale structures develop, as one also sees from
the energy spectra in \Fig{ppower_all_EK_EM_T1024eo5axi_powerbz}.

It should be pointed out that, owing to the persistent growth of
magnetic and kinetic energy, the Reynolds numbers grow eventually
beyond the limit of what can be resolved at a given resolution.
Unlike some of the earlier LES, where numerical effective viscosity and
diffusivity keep the small scales resolved, in the DNS of \cite{rogachevskiiea12}
this is not the case and the numerical code (in this case the
{\sc Pencil Code}\footnote{http://www.pencil-code.googlecode.com})
eventually `crashes'.
The point when this happens can be delayed by using higher resolution.
This is why we show here the results for ${\cal J}=800$ at a resolution
of $1024^3$ mesh points, where the simulation can be carried out
for about 0.126 Alfv\'en times, compared to only 0.09 Alfv\'en times
at a resolution of $512^3$ mesh points used in \cite{rogachevskiiea12}.
Remeshing the $1024^3$ run to $2048^3$ mesh points, we were able to
continue until 0.142 Alfv\'en times, after which we were unable to continue
the run due to a disk problem.

The Bell instability is driven by the simultaneous presence of
an external magnetic field $\BB_0$ and an external current $\jj_{\rm cr}$,
giving therefore rise to a pseudo-scalar $\jj_{\rm cr}\cdot\BB_0$;
here, $\BB_0$ is an axial vector while $\jj_{\rm cr}$ is a polar vector.
In stellar magnetism, the presence of a pseudo-scalar is caused by rotation
$\OO$ (an axial vector) and gravity $\grav$ (a polar vector).
This property is generally held responsible for the production of magnetic
fields by what is known as the $\alpha$ effect.
As explained in Sect.~4.4, the $\alpha$ effect denotes the presence
of a tensorial connection between a mean electromotive force
$\meanEMF=\overline{\uu\times\bb}$ and a mean magnetic field via
\begin{equation}
\meanemf_i=\alpha_{ij}\meanB_j+\eta_{ijk}\meanB_{j,k}+...,
\end{equation}
where higher order derivatives (indicated by commas) of the mean magnetic
field are also present.
If the tensors $\alpha_{ij}$ and $\eta_{ijk}$ were isotropic and the
evolution characterized by just two quantities,
$\alpha=\delta_{ij}\alpha_{ij}/3$ and $\etat=\epsilon_{ijk}\eta_{ijk}/6$,
the growth of the mean magnetic field would occur at the rate
\begin{equation}
\gamma_{\rm dynamo}=\alpha k-\etaT k^2,
\label{dynamogrowth}
\end{equation}
where $\etaT=\etat+\nuM$ is the total (turbulent plus microphysical)
magnetic diffusivity and the fastest growth occurs at wavenumber
$k=\alpha/2\etaT$ with the growth rate $\gamma_{\rm max}=\alpha^2/4\etaT$.

In stellar dynamos, where the magnetic Reynolds number is very large,
the actual growth is dominated by small-scale dynamo action, so
\Eq{dynamogrowth} is in practice not obeyed, unless the small-scale dynamo
is not excited, for example at low magnetic Prandtl numbers \citep{B09}.
However, in the present case the magnetic energy spectra show that
at late times, magnetic power moves gradually to larger scales.
This is why we now ask whether this can be explained by the
$\alpha$ effect.

\cite{rogachevskiiea12} have shown that in the case of $\jj_{\rm cr}$ and $\BB_0$
pointing in the $z$ direction, the large-scale mean magnetic field
is a function of $x$ and $y$ and can be written in terms of two scalar
functions $\meanA_\parallel(x,y,t)$ and $\meanB_\parallel(x,y,t)$ with
\begin{equation}
\meanBB(x,y,t)=\nab\times(\zzz\meanA_\parallel)+\zzz\meanB_\parallel,
\end{equation}
where $\zzz=(0,0,1)$ is the unit vector in the $z$ direction.
These functions obey the mean field equations
\begin{eqnarray}
\partial\meanA_\parallel/\partial t
=\alpha_A\meanB_\parallel+\eta_A\nabla^2\meanA_\parallel,
\label{EqA}
\\
\partial\meanB_\parallel/\partial t
=\alpha_B\meanJ_\parallel+\eta_B\nabla^2\meanB_\parallel,
\label{EqB}
\end{eqnarray}
where $\meanJ_\parallel=-\nabla^2\meanA_\parallel$ is the $xy$ dependent
part of the mean current density in the $z$ direction.
We consider a homogeneous system, so the coefficients $\alpha_A$,
$\alpha_B$, $\eta_A$, and $\eta_B$ are constant and we can seek solutions
of a form proportional to $\exp(\lambda t+\ii\kk\cdot\xx)$.
In this case, the dynamo growth rate is still described by
\Eq{dynamogrowth}, provided we substitute
\begin{equation}
\alpha\to\alpha^{\rm eff}=(\alpha_A\alpha_B+\epsilon_\eta^2k^2)^{1/2}
\quad\mbox{and}\quad
\etat\to\etat^{\rm eff}=(\eta_A+\eta_B)/2,
\label{substitute}
\end{equation}
where $\epsilon_\eta=(\eta_A-\eta_B)/2$ quantifies the anisotropy
of the turbulent diffusivity.

\begin{figure}
\begin{center}
\includegraphics[width=\columnwidth]{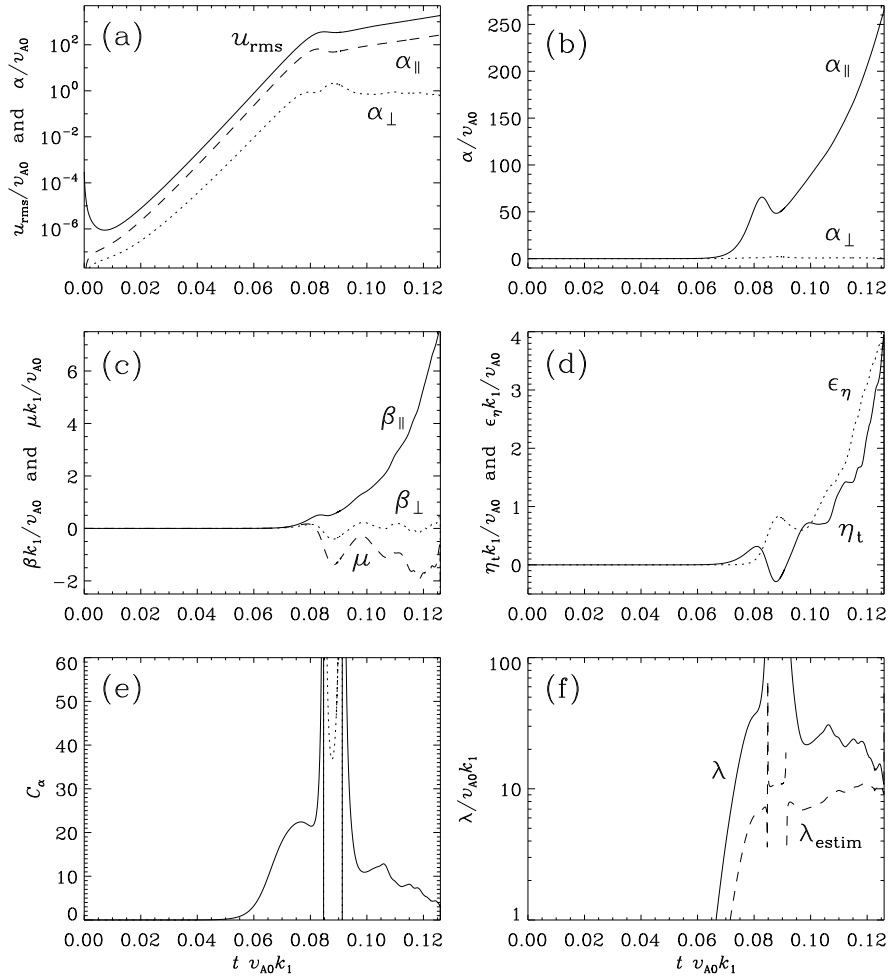}
\end{center}\caption[]{
Time evolution of the model parameters for ${\cal J}=800$ and $\Lu=80$
using $1024^3$ mesh points.
(a) Exponential growth and subsequent near-saturation of
$u_{\rm rms}$, $\alpha_\parallel$, and $\alpha_\perp$
(all normalized by $v_{\rm A0}$) in linear-logarithmic representation.
(b) Evolution of $\alpha_\parallel$ and $\alpha_\perp$
(normalized by $v_{\rm A0}$) in double linear representation,
showing that $\alpha_\perp$ is much smaller than $\alpha_\parallel$.
(c) Evolution of $\beta_\parallel$, $\beta_\perp$, and $\mu$
(normalized by $v_{\rm A0}/k_1$).
(d) Evolution of $\eta_{\rm t}$ and $\epsilon_\eta$
(normalized by $v_{\rm A0}/k_1$).
(e) Evolution of $C_\alpha$ (negative values are shown as dotted lines),
and (f) growth of the fasted growing mode.
}\label{palpaxi}
\end{figure}

To determine these coefficients from the DNS, we use the so-called
test-field method of \cite{Sch05}, which was originally used in spherical
coordinates.
The implementation in Cartesian coordinates is described in \cite{B05}
and especially in \cite{BRK12}, where the mean magnetic field was allowed
to depend on all three spatial coordinates, and not just on one, as was
assumed in \cite{B05}.
Under the assumption that the turbulence is governed by only one preferred
direction, which is here the case, the number of coefficients reduces to
9, and homogeneity reduces this number further to 5, so in the present
case we have
\begin{equation}
\meanEMF=\alpha_\perp\meanBB_\perp+\alpha_\parallel\meanBB_\parallel
-\beta_\perp\meanJJ_\perp+\beta_\parallel\meanJJ_\parallel
-\mu\zzz\times\meanKK_\perp,
\end{equation}
where $\meanJJ=\nab\times\meanBB$ characterizes the antisymmetric part
of the magnetic derivative tensor and
$\meanK_i=(\meanB_{i,j}+\meanB_{j,i})\hat{z}_j/2$ the symmetric part.
We have followed here the notation of \cite{BRK12}, except that there
the two $\alpha$ coefficients were defined with the opposite sign.
Comparing with the coefficients used in \Eqs{EqA}{EqB}, we find that
$\alpha_A=\alpha_\parallel$, $\alpha_B=\alpha_\perp$,
$\eta_A=\beta_\parallel$, and $\eta_B=\beta_\perp-\mu/2$.
In \Fig{palpaxi} we show the time dependence of the various parameter
combinations.
In the early kinematic phase ($t v_{\rm A0} k_1 < 0.08$), the root mean
square velocity, $u_{\rm rms}$, as well as $\alpha_\parallel$ and
$\alpha_\perp$ grow exponentially.
At later times, $\alpha_\parallel$ continues to grow, while $\alpha_\perp$
remains small and approximately constant.
The other turbulent transport coefficients also grow exponentially in the
kinematic phase, and at later times $\beta_\parallel$, $\eta_t$, and
$\epsilon_\eta$ continue to grow, while $\beta_\perp$ and $\mu$ remain
small and can even become negative.
The resulting effective dynamo number, which is proportional to the product
$\alpha_\parallel\alpha_\perp$, reaches values well above the critical value
of unity.
The estimated and actual growth rates agree roughly and have a value of
around 10 in units of $\eta_t k_1$.

\section{Instabilities driven by the nearly isotropic CR distributions}\label{drury}

In many astrophysical objects the CR mean free path due to the particle
scattering by magnetic fluctuations carried by the background
plasmas is below the characteristic scale sizes of the plasma flow.
In that case the angular distribution of the CRs is nearly isotropic
with a small anisotropic part (i.e. both $\beta \ll
1$ and $\chi \ll 1$ in Eq.(\ref{distrF0})). Then one can use the diffusion
approximation that assumes
\begin{equation}\label{F_crDiff}
f^{cr}(\mathbf{r},\mathbf{p})=\frac{1}{4\pi}\left[N^{cr}(\mathbf{r},p)+
\frac{3}{v\,p}\mathbf{p}\mathbf{J}^{cr}(\mathbf{r},p)\right],
\end{equation}
where the diffusive current of CRs is
\begin{equation}\label{JcrDiff}
J_{\alpha}^{cr}=-\kappa_{\alpha\beta}\nabla_{\beta}N^{cr}-\frac{p}{3}\frac{\partial
N^{cr}}{\partial p}u_{\alpha},
\end{equation}
$\kappa_{\alpha\beta}$ is the  momentum-dependent CR diffusion
tensor. Then the kinetic equation Eq.(\ref{KinEqCR1}) reduces to the
advection-diffusion equation for the isotropic part of CR
distribution $N^{cr}(\mathbf{r},p,t)$
\begin{equation}\label{EqDiffConv}
\frac{\partial N^{cr}}{\partial
t}=\nabla_{\alpha}\kappa_{\alpha\beta}\nabla_{\beta}N^{cr}-
\left(\mathbf{u}\nabla\right)N^{cr}+\frac{p}{3}\frac{\partial
N^{cr}}{\partial p}\nabla\mathbf{u},
\end{equation}
where $\mathbf{u}(\mathbf{r},t)$ is the bulk velocity of the
background plasma \citep[see, e.g.,][]{Toptygin83}. It is assumed
here for simplicity that the scatterers are carried with the plasma
bulk velocity, though it is possible to account for the scatterers
drift velocity \citep[see, e.g.,][]{skilling75}. The advantage of
this approach is that it is valid for collision operators $I[f]$
more general than just the simple relaxation time approximation
given by Eq.(\ref{integrSt2}). In the diffusion approximation the
exact form of the collision operator determines the form of the
diffusion tensor and its momentum dependence. Therefore,  the
results obtained within the diffusion approximation are valid for
different collision operators.

To explore the effect of CRs on the background plasma one should
calculate the first moment of the kinetic equation
Eq.(\ref{KinEqCR1}) for CRs that is the momentum exchange rate
between the CRs and the background plasma:
\begin{equation}\label{intKinCr}
\frac{\partial \mathcal{P}_{\alpha}}{\partial
 t}+\nabla_{\alpha} P^{cr} +\nabla_{\beta} \Pi'_{\alpha\beta}=
\left[\frac{1}{c}(\mathbf{j}^{cr}-en_{\rm
cr}\mathbf{u})\times\mathbf{B}+\int
\mathbf{p}I[f]d^{3}p\right]_{\alpha},
\end{equation}
where $P^{cr}$ is the CR pressure, the CR momentum density
\begin{equation}\label{38d}
  \mathbf{\mathcal{P}}(\mathbf{r},\ t)=\int\mathbf{p}f\,d^{\,3}p ,
\end{equation}
and the reduced CR momentum flux density  $\Pi'_{\alpha\beta}$ is
defined by
\begin{equation}\label{e}
 \Pi'_{\alpha\beta}=\int p_\alpha v_\beta f\,d^{\,3}p -
 P^{cr}\delta_{\alpha\beta}.
\end{equation}

In the diffusion approximation for the steady state (e.g., in the
shock rest frame)  the first and the third terms in the left hand
side of Eq.(\ref{intKinCr}) are small and then
Eq.(\ref{eqMotiontot}) can be reduced to
\begin{equation}\label{eqMotiontotLarg}
\rho\left(\frac{\partial\mathbf{u}}{\partial
 t}+(\mathbf{u}\nabla)\mathbf{u}\right)
 =- \nabla \left(p_{g}+P^{cr}\right)+
 \frac{1}{4\pi}(\nabla\times\mathbf{B})\times\mathbf{B}.
\end{equation}
The equation can be applied to longwavelength perturbations. It
should be supplied with the continuity equation:
\begin{equation}\label{Ch1}
\frac{\partial \rho}{\partial
t}+\nabla\left(\rho\mathbf{u}\right)=0,
\end{equation}
the energy equations for the background plasma:
\begin{equation}\label{Ch3}
 \frac{\partial p_{g}}{\partial t}+\left(\mathbf{u}\nabla\right)p_{g}+\gamma_{g}p_{g}\nabla\mathrm{u}=0,
\end{equation}
the MHD induction equation
\begin{equation}\label{Ch2}
\frac{\partial\mathbf{B}}{\partial t}=
\nabla\times\left(\mathbf{u}\times\mathbf{B}\right),~~~
\nabla\mathbf{B}=0,
\end{equation}
and the equation for CR-pressure variations
 \begin{equation}\label{Ch4}
 \frac{\partial P^{cr}}{\partial t}+\left(\mathbf{u}\nabla\right)P^{cr}+\gamma_{cr}P^{cr}\nabla\mathrm{u}=\nabla_{\alpha}\overline{\kappa}_{\alpha\beta}\nabla_{\beta}P^{cr},
\end{equation}
where  $\overline{\kappa}_{\alpha\beta}$  is the CR diffusion tensor
averaged over the CR distribution function, $\gamma_{g}$ and
$\gamma_{cr}$ - are the adiabatic indexes of the plasma and CRs,
respectively.

\section{Acoustic instability  driven by the CR pressure
gradient}\label{drury1}

It was found by \citet{Drury84,dorfi85,DruryFal86,DruryDrInst12}
that the force density in Eq.(\ref{eqMotiontotLarg}) associated with
the CR pressure gradient that does not depend on the density of the
background plasma results in a specific instability. The effect of
magnetic field on the instability was studied by \citet{berezhko86}
and \citet{Chalov88}. The analytical study of the instability can be
performed for the modes with the wavenumbers below the scale size of
the CR pressure gradient $L \sim P^{cr}/|\nabla P^{cr}|$. In the
generic case of the diffusive shock acceleration $L \sim (c/u_{\rm
s})\times r_{g}/a$. Following \citet{DruryFal86,Chalov88} for the
wavenumber range $kL >1$, but $kr_{g}/a < 1$ the mode growth and
damping can be derived from the continuity equation for the wave
action.

The mode growth rate $\Gamma$ in the simplified geometry where the
CR pressure gradient is directed along the unperturbed magnetic
field was derived using a standard linear analysis of
Eqs.(\ref{eqMotiontotLarg} -- \ref{Ch4}) by \cite{Chalov88a}, who
obtained the following expression
\begin{eqnarray}\label{GamChalov}
& &
\Gamma=\frac{v_{m}^{2}-v_{a}^{2}}{2v_{m}^{2}-(v_{s}^{2}+v_{a}^{2})}\left\{-\frac{\gamma_{cr}P^{cr}_0}{\rho_{0}}\frac{k^{2}}{\kappa_{0\parallel}k_{\parallel}^{2}+\kappa_{0\perp}k_{\perp}^{2}}
\frac{v_{m}^{2}-v_{a}^{2}\frac{k_{\parallel}^{2}}{k^{2}}}{v_{m}^{2}-v_{a}^{2}}\pm\right.
 \nonumber\\
 & &
\left.\pm\frac{\nabla
P^{cr}_0}{\rho_{0}v_{m}}\frac{k_{\parallel}}{k}\left[1+\frac{\varsigma\kappa_{0\parallel}k^{2}}{\kappa_{0\parallel}k_{\parallel}^{2}+\kappa_{0\perp}k_{\perp}^{2}}\frac{v_{m}^{2}-v_{a}^{2}\frac{k_{\parallel}^{2}}{k^{2}}}{v_{m}^{2}-v_{a}^{2}}\right]\right\}.
\end{eqnarray}
Here $v_{s}$  is the sound speed of the background plasma,
$P^{cr}_0$  is the unperturbed CR pressure, $\nabla P^{cr}_0$ is the
gradient of the unperturbed CR pressure, $k_{\parallel}$ and
$k_{\perp}$  are the components of the mode wavevector parallel and
transverse to the unperturbed magnetic field, respectively, and
$\kappa_{0\parallel}$, $\kappa_{0\perp}$ are the components of the
averaged CR diffusion tensor. It is assumed that the CR diffusion
tensor components scale with the background plasma density as
$\overline{\kappa}_{\parallel,\perp}\sim \rho^{\varsigma}$. The
phase velocity of the mode is
\begin{equation}\label{vph}
v_{m}=\left[v_{s}^{2}+v_{a}^{2}\pm\frac{1}{2}\sqrt{\left(v_{s}^{2}+v_{a}^{2}\right)^{2}-4v_{s}^{2}v_{a}^{2}\frac{k_{\parallel}^{2}}{k^{2}}}\right]^{\frac{1}{2}}.
\end{equation}
The first term in Eq.(\ref{GamChalov}) is the wave damping rate due
to the irreversible stochastic Fermi II CR acceleration effect
\citep{achterberg79,bt79,Ptuskin81}, while the second and the third
terms describe the growth/damping of the modes due to the acoustic
instability studied by \citet{DruryFal86}. A more general treatment
with an arbitrary direction of the unperturbed magnetic field was
performed by \cite{Chalov88}. He accounted for the response of the
CR diffusion tensor to both the density and  magnetic field
variations and found that the latter does not change the character
of the angular dependence of the growth rate significantly. A
similar angular dependence of the long-wave mode growth rate due to
the CR current driven instability (discussed above in \S\ref{LW})
was found by \citet{Bykov11}.

In the space plasma with the modest level of the magnetic field
fluctuations the local CR diffusion is anisotropic. For magnetized
CR particles ($a\ll 1$) the diffusion parallel to the mean magnetic
field dominates over the CR diffusion transverse to the mean field,
i.e., $\kappa_{0\parallel}\gg\kappa_{0\perp}$. The growth rate of
the acoustic instability in the anisotropic system is maximal for
the modes propagating nearly transverse to the mean magnetic field
($\vartheta \rightarrow \displaystyle\frac{\pi}{2}$). Here
$\vartheta$ is the angle between the mode wavevector and the the
mean magnetic field.

\begin{figure*}
\centering
\includegraphics[width=10cm]{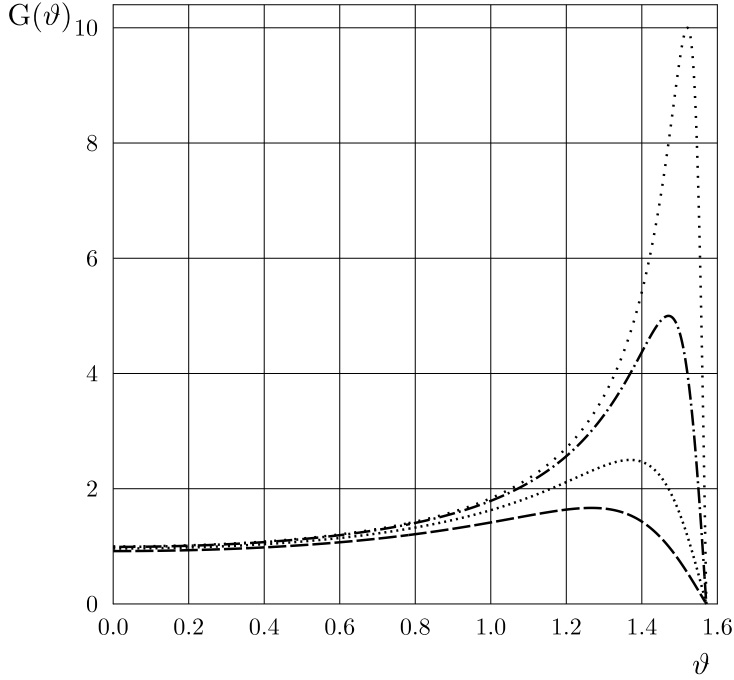}
\caption{The characteristic angular dependence of the growth rate of
the acoustic instability Eq.(\ref{GamChalov}) for $a$ =0.3  (the
dashed curve), $a$=0.2  (the dotted curve), $a$=0.1  (the dot-dashed
curve), $a$=0.05 (the rare dot curve).} \label{figAngelCoef}
\end{figure*}

The angular dependence of the growth rate Eq.(\ref{GamChalov}) can
be approximated by
\begin{equation}\label{AngelCoef0}
G_{0}(\vartheta)=\frac{\cos\vartheta}{\cos^{2}\vartheta+\frac{\kappa_{0\perp}}{\kappa_{0\parallel}}\sin^{2}\vartheta},
\end{equation}
where we used $\frac{k_{\parallel}}{k}=\displaystyle\cos\vartheta$,
$\frac{k_{\perp}}{k}=\displaystyle\sin\vartheta$. The anisotropy of
the CR diffusion is determined by the CR particle magnetization
\citep[e.g.,][]{Toptygin83}, that is the inverse collisionality
parameter $a$, and therefore,
$\displaystyle\frac{\kappa_{0\perp}}{\kappa_{0\parallel}}\propto
a^{2}$. The maximal growth rate is therefore achieved for the mode
propagating at $\cos\vartheta_{\rm max}=a$, where $G_{\rm
max}(\vartheta_{\rm max})=\displaystyle\frac{1}{2a}$. The angular
dependence of the growth rate of the acoustic instability is
illustrated in Figure \ref{figAngelCoef} for various values of
collisionality parameter $a$.

The linear perturbation analysis discussed above is based on the
diffusion approximation of the CR dynamics in
Eqs.(\ref{eqMotiontotLarg} - \ref{Ch4}) and, therefore, it is valid
for the modes of the wavenumbers above the mean free path of the
CRs. A numerical model of the acoustic instability in the nonlinear
regime was performed recently by \citet{DruryDrInst12}, who found a
significant amplification of magnetic field. The authors assumed a
fixed CR diffusion gradient with no response of the CR pressure
to the fluctuations, that may affect the model results.

\section{Self-confinement of CRs near their acceleration sites}

Apart from being a central issue for the acceleration in SNR shocks,
the CR-driven instabilities are fast becoming an integral part of CR
escape models. One common difficulty with the observational
verification of the proton escape is that, in contrast to electrons,
they likely remain invisible until they reach some dense material in
SNR surroundings. Only there generate they enough $\pi^{0}$ mesons
in collisions with other protons and the mesons in turn decay into
gamma photons which may be detected. Not surprisingly, the escape of
CRs from an SNR is a hot topic of today research in gamma-ray
astronomy.

The backbone of the DSA is a self-confinement of accelerated
particles by scattering off various magnetic perturbations that
particles drive by themselves while streaming ahead of the shock.
Most important of them were discussed at some length in this review.
Logically, this process should also control the ensuing propagation
of CRs, before their density drops below the wave instability
threshold. Strictly speaking the CR release (escape) from the
accelerator should be treated together with the acceleration, as it
does not occur at once for all the particles. But this would be a
combination of two difficult enough problems and most of the
progress in CR escape was made by considering it separately from
acceleration.

Remarkably, even within this limited approach, and under rather
loose formulation of the problem, no consensus on the escape mechanism
 has been reached so
far; the dividing lines seem to run across the following issues: (i)
does the escape occur isotropically or along the local magnetic
field? (ii) does the scattering by the background MHD turbulence
control the CR propagation alone or self-excited waves need to be
included? (iii) if so, is a quasilinear saturation of self-excited
waves sufficient or nonlinear processes of wave damping are crucial
to the particle propagation? (iv) if they are, which particular
mechanism(s) should be employed?

Starting with (i-ii) we note that most of the early models, and some
of the recent ones that target specific remnants, assume isotropic
CR propagation from a point source impeded only by the background
turbulence (one may call them test particle models, e.g.
\citealp{AharAt96,GabiciAharEsc09,eb11,Gabici11}). It should be
noted, however, that e.g., \citet{Rosner96} and
\citet{GabiciAnisEsc12} adopted a field aligned propagation while
\citet{DruryEscape11} included the finite radius of a SNR shock in
the CR escape description. Given the topic of the present short
review, however, we focus in this section on models that explicitly
include the self-excited waves. Brief reviews of other aspects of CR
propagation in the galaxy were given recently by \citet{Gabici11}
and \citet{PtuskinPropRev12}.

The role of self-confinement effects in the CR escape, their
subsequent propagation and how these phenomena are treated in
different models, can be best demonstrated by writing the following
equations that self-consistently describe the CR diffusion and wave
generation

\begin{equation}
\frac{d}{dt}P_{{\rm {\rm
CR}}}\left(p\right)=\frac{\partial}{\partial z}\frac{\kappa_{{\rm
B}}}{I}\frac{\partial P_{{\rm CR}}}{\partial z}\label{eq:dPdt}
\end{equation}

\begin{equation}
\frac{d}{dt}I=-v_{{\rm a}}\frac{\partial P_{{\rm CR}}}{\partial
z}-\Gamma I.\label{eq:dIdt}
\end{equation}
Here $v_{a}$ is the Alfv\'en velocity, $\kappa_{B}$ is the CR
diffusion coefficient in Bohm regime, $\kappa_{B}=cr_{g}/3$, and the
time derivative is taken along the characteristics of unstable
Alfv\'en waves, forward propagating along the field ($z$-
direction):

\begin{equation}
\frac{d}{dt}=\frac{\partial}{\partial t}+v_{{\rm
a}}\frac{\partial}{\partial z}\label{eq:charct}
\end{equation}
Eq.(\ref{eq:dPdt}) above is essentially a well-known
convection-diffusion equation, written for the dimensionless CR
partial pressure $P_{{\rm CR}}$ instead of their distribution
function $f\left(p,t\right)$. We normalized it to the magnetic
energy density $\rho v_{{\rm a}}^{2}/2$:

\begin{equation}
P_{{\rm CR}}=\frac{4\pi}{3}\frac{2}{\rho v_{{\rm
a}}^{2}}vp^{4}f,\label{eq:PcrDef}
\end{equation}
where $v$ and $p$ are the CR speed and momentum, and $\rho$- the
plasma density. The total CR pressure is normalized to $d\ln p$,
similarly to the wave energy density $I$:

\[
\frac{\left\langle \delta B^{2}\right\rangle
}{8\pi}=\frac{B_{0}^{2}}{8\pi}\int I\left(k\right)d\ln
k=\frac{B_{0}^{2}}{8\pi}\int I\left(p\right)d\ln p
\]
Eq.(\ref{eq:dIdt}) is a wave kinetic equation in which the energy
transferred to the waves equals the total
work done by the particles, $\left(u+v_{{\rm a}}\right)\nabla
P_{{\rm CR}}$, less the work done on the fluid, $u\nabla P_{{\rm
CR}}$ \citep{Drury83} (we neglect the bulk flow velocity $u$, here
and in Eq.(\ref{eq:charct}) assuming that the active phase of
acceleration ended by this time). The above interpretation of the
wave generation indicates that it operates in a maximum efficiency
regime. A formal quasilinear derivation of this equation assumes
that the particle momentum $p$ is related to the wave number $k$ by
the 'sharpened' resonance condition $kp=eB_{0}/c$ instead of the
conventional cyclotron resonance condition $kp_{\|}=eB_{0}/c$
\citep{skilling75}, (note that here $k=k_{\parallel})$. We assume that
$\partial P_{{\rm CR}}/\partial z\le0$ at all times, so that only
the forward propagating waves are unstable. The latter inequality is
ensured by the formulation of initial value problem symmetric with
respect to $z=0$, so we consider the CR escape into the
half-space $z>0$ with the boundary condition $\partial P_{{\rm
CR}}/\partial z=0$ at $z=0$.

Papers on CR self-confinement discussed below use equations that are
largely similar to Eqs.(\ref{eq:dPdt}-\ref{eq:dIdt}) but different
assumptions are made regarding geometry of particle escape from the
source (see (i) above), the character and strength of wave
damping $\Gamma$ (iv), and the role of quasilinear wave saturation
(iii). \citet{Fujita11} and \citet{YanLazarianEscape12} utilize the
isotropic escape models (in this case $\partial/\partial z$ should
be replaced by $\partial/\partial r$, etc.) while
\citet{PtuskinNLDIFF08} and \citet{Malkov_etal_escape12} assume that
particles propagate predominantly along the local large-scale field.
Note that \citet{YanLazarianEscape12} considered the escape from an
 active accelerator (in Eq.(\ref{eq:charct}), one should
include the flow bulk velocity, $v_{a}\to v_{a}+u$ in this case)
and, in addition, they introduce a step-wise increase in CR
diffusivity at a certain particle momentum above which
particles escape the accelerator. These assumptions make it
difficult to compare their results with those of the remaining three
papers. In these, \citet{Fujita11} presented the results of
numerical integration of Eqs.(\ref{eq:dPdt}-\ref{eq:dIdt}) (in a
spherical symmetry) with neglected damping term $\Gamma$. The
results indicate a considerable delay of diffusion from the source
due to a self-confinement.

\begin{figure}
\includegraphics[scale=0.5, angle=-90]{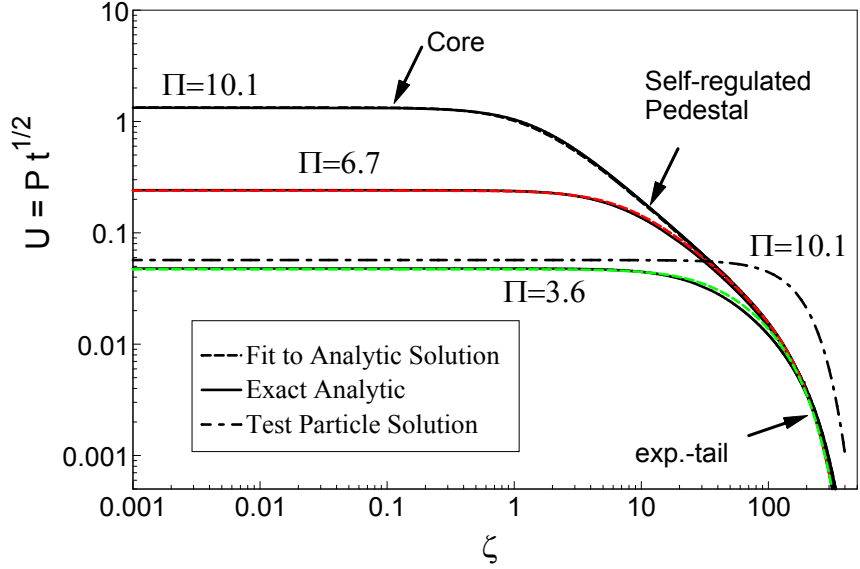}
\caption{Spatial distribution of CR partial pressure (as a function
of $\zeta=z/\sqrt{av_{a}t}$, multiplied by
$v_{a}^{3/2}\sqrt{at}/\kappa_{B}$) shown for integrated values of
this quantity $\Pi=3.6;$ 6.7; 10.1 and for for the background
diffusivity $D_{ISM}=10^{4}.$ Exact analytic solutions are shown
with the solid lines while the interpolations given by Equation
(\ref{eq:Ufit}) are shown with the dashed lines. For comparison, a
formal test particle solution for $\Pi=10.1$ is also shown with the
dot-dashed line. Note the three characteristic zones of the CR
confinement: the innermost flat top core, the scale invariant
($1/\zeta$) pedestal, and the exponential decay zone.
\label{fig:SpatialProfileFits}}
\end{figure}

However, in the regions where magnetic perturbations are weak, i.e.
$I\ll1$, the field aligned CR transport is appropriate, as the
perpendicular diffusion is suppressed, $\kappa_{\perp}\simeq
I^{2}\kappa_{\parallel}\ll\kappa_{\parallel}\simeq\kappa_{B}/I$.
Taking into account the condition $I_{{\rm ISM}}\ll1$, such regime
appears inevitable far away from the source and at late times when
particles are spread over a large volume and the waves are driven
only weakly. At earlier times and close to the region of the initial
localization of CRs, an estimate
$\kappa_{\perp}\sim\kappa_{\parallel}\sim\kappa_{{\rm B}}$ appears
to be adequate. Both analytical models by \citet{PtuskinNLDIFF08}
and \citet{Malkov_etal_escape12}, however, do not embrace the
general case and rely on the assumption
$\kappa_{\perp}\ll\kappa_{\parallel}$ thus considering a
field-aligned escape. At the same time, they are different in
further simplifications made, that lead to rather different results,
both quantitatively and qualitatively.

\citet{PtuskinNLDIFF08} neglect $dI/dt$ on the l.h.s of
Eq.(\ref{eq:dIdt}) thus balancing the driving term with the damping
term on its r.h.s and assume a Kolmogorov dissipation for $\Gamma$,
\begin{equation}
\Gamma=kv_{a}\sqrt{I}/\left(2C_{K}\right)^{3/2}\label{eq:GammaPZ}
\end{equation}
with $C_{K}\approx3.6$ and $k\simeq 1/r_{g}\left(p\right)$ being the
resonant wave number. Therefore, only one equation (\ref{eq:dPdt})
needs to be solved which lead to the following self-similar solution (in
notations and normalization used in
Eqs.(\ref{eq:dPdt}-\ref{eq:dIdt}))

\begin{equation}
P_{CR}=\frac{4\cdot3^{-3/2}}{t^{\prime3/2}\sqrt{\sigma+\left(kz\right)^{4}/t^{\prime6}}}\label{eq:PressurePZ}
\end{equation}
where the dimensionless time
$t^{\prime}=\left(\kappa_{B}k^{2}/2C_{K}\right)t$,
$\sigma=\Gamma^{8}\left(1/4\right)/\pi^{2}3^{6}\eta^{4}$, and
$\Gamma $ is the gamma function. The single important parameter this
solution depends on is the integrated (along the field line) CR
partial pressure
\begin{equation}
\eta=2k\intop_{0}^{\infty}P_{CR}dz\label{eq:IntPressure}
\end{equation}
Therefore, the CR density decays at the source as $\propto t^{-3/2}$
and the flat-topped, self-confined part of the CR distribution
spreads as $z\propto t^{3/2}$, both pointing at the superdiffusive
CR transport. The reason is clearly in a very strong wave damping
due to the Kolmogorov dissipation. For the same reason this solution
does not recover the test particle asymptotic result
$P_{CR}\propto t^{-1/2}\exp\left(-z^{2}/4D_{{\rm ISM}}t\right)$,
physically expected in $z,t\to\infty$ limit in the interstellar
medium with the background diffusion coefficient $D_{{\rm ISM}}$.

An alternative choice of damping mechanism is the \citet{goldr97}
MHD spectrum, which seems to be more appropriate in $I\lsim1$ regime
under not too strong MHD cascade
\citep{FarmerGoldr04,BeresnLaz08,YanLazarianEscape12}. The damping
rate in this case is
\begin{equation}
\Gamma=v_{a}\sqrt{\frac{k}{L}}\label{eq:GammaGS}
\end{equation}
where $L$ is the outer scale of turbulence which may be as large as
$100pc$. Not only is this damping orders of magnitude (roughly a
factor $\sqrt{r_{g}/L}$ ) lower than the Kolmogorov one but, as it
does not depend on $I$ and can be considered as coordinate
independent, it allows for the following ('quasilinear') integral of
the system of Eqs.(\ref{eq:dPdt}) and (\ref{eq:dIdt}):

\begin{equation}
P_{{\rm CR}}\left(z,t\right)=P_{{\rm
CR}0}\left(z^{\prime}\right)-\frac{\kappa_{{\rm B}}}{v_{{\rm
a}}}\frac{\partial}{\partial
z}\ln\frac{I\left(z,t\right)}{I_{0}\left(z^{\prime}\right)}\label{eq:QLint}
\end{equation}
Here $P_{{\rm CR}0}\left(z\right)$ and $I_{0}\left(z\right)$ are the
initial distributions of the CR partial pressure and the wave energy
density, respectively, and $z^{\prime}=z-v_{{\rm a}}t$. Substituting
$P_{CR}$ in Eq.(\ref{eq:dIdt}) and neglecting slow convection with
$v_{a}$ in Eq.(\ref{eq:charct}), we arrive at the following
diffusion equation for $I$
\[
\frac{\partial I}{\partial t}=\frac{\partial}{\partial
z}\frac{\kappa_{B}}{I}\frac{\partial I}{\partial z}-\Gamma
I-v_{a}\frac{\partial P_{CR0}}{\partial z}.
\]
The equation is supplemented with the boundary condition   $I\to
I_{ISM}$, for $|z|\to
 \infty$. Outside of the region where $P_{CR}\neq0,$ the last term on the
r.h.s. may be neglected. The second term may be eliminated by
replacing $I\exp\left(\Gamma t\right)\to I$,
$\intop_{0}^{t}\exp\left(\Gamma t\right)dt\to t$. However, if
$\Gamma$ is taken in the form of Eq.(\ref{eq:GammaGS}), it is fairly
small due to the factor $\sqrt{r_{g}/L}\ll1$. We may simply neglect
it. The solution for $I$ and $P_{CR}\left(z,t\right)$ may be found
in an implicit form (see \citealp{Malkov_etal_escape12} for
details). However, there exists a very accurate convenient
interpolation formula that can be represented as follows

\begin{equation}
P_{{\rm CR}}=\frac{2\kappa_{{\rm B}}\left(p\right)}{v_{{\rm
a}}^{3/2}\sqrt{L_{c}t}}\left[\zeta^{5/3}+\left(D_{{\rm
NL}}\right)^{5/6}\right]^{-3/5}e^{-\zeta^{2}/4D_{{\rm
ISM}}}\label{eq:Ufit}
\end{equation}
where $L_{c}$ is the size of the initial CR cloud,
$\zeta=z/\sqrt{v_{a}L_{c}t}$, and $D_{{\rm
NL}}=C\left(\Pi\right)D_{{\rm ISM}}\exp\left(-\Pi\right)$, with
$\Pi$ being a normalized integrated pressure

\[
\Pi=\frac{v_{a}}{\kappa_{B}}\intop_{0}^{\infty}P_{CR}dz
\]
and $D_{{\rm ISM}}$ is the normalized background diffusivity

\[
D_{{\rm ISM}}=\frac{\kappa_{B}}{v_{a}L_{c}}I_{{\rm ISM}}^{-1}
\]
while $C\sim1$, for $\Pi\gg1$ and $C\sim{\Pi}^{-2}$, for $\Pi\ll1$.

The representation of the solution given in Eq.(\ref{eq:Ufit}) is
convenient in that the function $\sqrt{t}P_{CR}\left(\zeta\right)$
does not depend on $t$, so that the solution can be shown for all
$t,z$ with only one curve, Figure \ref{fig:SpatialProfileFits}. To
summarize these results, the self-regulated normalized
($\mathcal{P}_{CR}=v_{{\rm a}}L_{c}P_{{\rm CR}}/\kappa_{{\rm
B}}\left(p\right)$) CR partial pressure profile $\mathcal{P}_{CR}$
comprises the following three zones ($\Pi\gg1$): (i) a quasi-plateau
(core) at small $z/\sqrt{t}<\sqrt{D_{{\rm NL}}}$ of a height
$\sim\left(D_{{\rm NL}}t\right)^{-1/2}$, which is elevated by a
factor $\sim\Pi^{-1}\exp\left(\Pi/2\right)\gg1$, compared to the
test particle solution because of the strong quasi-linear
suppression of the CR diffusion coefficient with respect to its
background (test particle) value $D_{{\rm ISM}}$: $D_{{\rm NL}}\sim
D_{{\rm ISM}}\exp\left(-\Pi\right)$ (ii) next to the core, where
$\sqrt{D_{{\rm NL}}}<z/\sqrt{t}<\sqrt{D_{{\rm ISM}}}$, the profile
is scale invariant, $\mathcal{P}_{CR}\approx2/z$. The CR
distribution in this ``pedestal'' region is fully self-regulated,
independent of $\Pi$ and $D_{{\rm ISM}}$ for $\Pi\gg1$, (iii) the
tail of the distribution at $z/\sqrt{t}>\sqrt{D_{{\rm ISM}}}$ is
similar \emph{in shape }to the test particle solution in 1D but it
saturates with $\Pi\gg1$, so that the CR partial pressure is
$\propto\left(D_{{\rm ISM}}t\right)^{-1/2}\exp\left(-z^{2}/4D_{{\rm
ISM}}t\right)$, independent of the strength of the CR source $\Pi$,
in contrast to the test-particle result which scales as
$\propto\Pi$. Because of the CR diffusivity reduction, the CR cloud
half-life is increased and the cloud width is decreased, compared to
the test particle solution.

Depending on the functions $\Pi\left(p\right)$ and $D_{{\rm
ISM}}\left(p\right)$, the resulting CR spectrum generally develops a
spectral break for the fixed values of $z$ and $t$ such that
$z^{2}/t\sim D_{{\rm NL}}\left(p\right)\sim D_{{\rm
ISM}}\exp\left(-\Pi\right)$.

\section{Summary}
Cosmic rays, being a highly non-equilibrium component, often
comprise an energy density that is comparable to the ram pressure of
energetic plasma flows and magnetic fields in astrophysical sources
with high energy release such as supernova remnants, fast stellar
winds, and astrophysical jets of different scales. CRs may also play
a role in the global dynamics of interstellar gas in galaxies, in
particular, they may support galactic winds. In the presence of
gravitation, the buoyancy of CRs and magnetic field at galactic
scales may result in the magnetic Parker instability
\citep[][]{parker66,parker67,shu74,ryuea03,hanaszea09}. The local CR
diffusion is an important factor for the Parker instability to
occur.

The microphysical instabilities discussed above lay the groundwork
for detailed simulations of the global interstellar matter dynamics.
In this review we addressed the recent progress in understanding of
the CR-driven instabilities with special attention to
non-relativistic shocks. We started with a quasi-linear analysis of
the growth rates of the instabilities driven by anisotropic and
inhomogeneous CR distributions. Time dependent nonlinear simulations
are needed to draw conclusions about the saturation level and the
spectra of magnetic fluctuations produced by the non-equilibrium CR
distributions. We used numerical simulations to illustrate the
nonlinear dynamics of magnetic fluctuations. The CR-driven
instabilities are shown to be crucial for modeling particle
acceleration sources and the CR escape from the sources into the
interstellar matter.

\begin{acknowledgements}
We thank Andre Balogh and the ISSI staff for providing an inspiring
atmosphere at the International Space Science Institute Workshop in
Bern in 2012, which has led to new collaborations and scientific
progress. Computing resources to A.B. were provided by the Swedish
National Allocations Committee at the Center for Parallel Computers
at the Royal Institute of Technology in Stockholm and the High
Performance Computing Center North in Ume{\aa}. A.B. was supported
in part by the European Research Council under the AstroDyn Research
Project No.\ 227952 and the Swedish Research Council under the
project grant 621-2011-5076. A.M.B. and S.M.O. acknowledge support
from the RAS Programs P21 and OFN 16, and from the Ministry of
Education and Science of Russian Federation (Agreement No. 8409,
2012). Some of the simulations were performed at the Joint
Supercomputing Centre (JSCC RAS) and the Supercomputing Centre at
Ioffe Institute, St.Petersburg. M.M. acknowledges support by the
Department of Energy, Grant No. DE-FG02-04ER54738.
\end{acknowledgements}


\bibliographystyle{svjour}
\bibliography{bibliogr}



\section{Appendix A}

The dispersion equation Eq.(\ref{dispers}) can be expressed in the
elementary functions by evaluating  Eq. (\ref{sigma0Int}) and
Eq.(\ref{sigma1Int}) as
\begin{eqnarray}\label{sigma0}
& &
\sigma_{0}(p)=\frac{3}{2x^{2}}+\frac{3}{8x}\left(1-\frac{1}{x^{2}}+\left(\frac{a}{x}\right)^{2}\right)\Psi_{1}
-\frac{3a}{2x^{3}}\Psi_{2}\mp
 \nonumber \\
 &\mp& i\left\{\frac{3}{4x}\left(1-\frac{1}{x^{2}}+\left(\frac{a}{x}\right)^{2}\right)\Psi_{2}
 -\frac{3a}{2x^{2}}+\frac{3a}{4x^{3}}\Psi_{1}\right\},
\end{eqnarray}
\begin{eqnarray}\label{sigma1}
& &
\sigma_{1}(p)=\mp\frac{1}{x}\pm\frac{3}{2x^{3}}\mp\frac{3}{2x}\left(\frac{a}{x}\right)^{2}\pm\frac{3}{8x^{2}}\left(1-\frac{1}{x^{2}}+3\left(\frac{a}{x}\right)^{2}\right)\Psi_{1}\pm
 \nonumber \\
 &\pm&
\frac{3a}{4x^{2}}\left(1-\frac{3}{x^{2}}+\left(\frac{a}{x}\right)^{2}\right)\Psi_{2}-
 \nonumber \\
 &-&
 i\left\{\frac{3}{4x^{2}}\left(1-\frac{1}{x^{2}}+3\left(\frac{a}{x}\right)^{2}\right)\Psi_{2}
 -\frac{3a}{x^{3}}-\frac{3a}{8x^{2}}\left(1+\left(\frac{a}{x}\right)^{2}-\frac{3}{x^{2}}\right)\Psi_{1}\right\},
\end{eqnarray}
\begin{equation}\label{Psi 1}
\Psi_{1}(x)=ln\left[\frac{(x+1)^{2}+a^{2}}{(x-1)^{2}+a^{2}}\right],
\end{equation}
\begin{equation}\label{Psi 2}
\Psi_{2}(x)=arctg\left(\frac{x+1}{a}\right)+arctg\left(\frac{x-1}{a}\right),
\end{equation}

\end{document}